\documentclass{aastex631}

\usepackage{amsmath}
\usepackage{empheq}
\usepackage{mathrsfs}
\usepackage{textcomp}
\usepackage{enumitem}   
\usepackage{gensymb}
\usepackage{hyperref}
\usepackage{graphicx}
\usepackage{xcolor}
\usepackage[caption=false]{subfig}
\usepackage{multirow}
\usepackage{longtable}
\usepackage[flushleft]{threeparttable}
\usepackage{booktabs} 
\usepackage{CJK}
\usepackage[T1]{fontenc} 
\hypersetup{colorlinks, linkcolor={blue}, citecolor={blue}, urlcolor={blue}} 

\definecolor{DarkOrange}{RGB}{204, 85, 0}
\definecolor{LincolnGreen}{RGB}{17, 102, 0}
\def\ion#1#2{#1$\;${\footnotesize\rm{#2}}\relax}

\usepackage{xspace}

\newcommand\nicer{\textit{NICER}\xspace}

\newcommand\swift{\textit{Swift}\xspace}
\newcommand\xmm{\textit{XMM-Newton}\xspace}

\newcommand\rosat{\textit{ROSAT}\xspace}

\newcommand{\sersic}{{S\'{e}rsic}}

\newcommand\cps{$\rm count\,s^{-1}$\xspace}

\begin{document}

\title{Sub-relativistic Outflow and Hours-Timescale 
Large-amplitude X-ray Dips during 
Super-Eddington Accretion onto a Low-mass Massive Black Hole in the Tidal Disruption Event AT2022lri}

\author[0000-0001-6747-8509]{Yuhan Yao}\email{yuhanyao@berkeley.edu}
\affiliation{Miller Institute for Basic Research in Science, 468 Donner Lab, Berkeley, CA 94720, USA}
\affiliation{Department of Astronomy, University of California, Berkeley, CA 94720, USA}

\author[0000-0002-5063-0751]{Muryel Guolo}
\affiliation{Department of Physics and Astronomy, Johns Hopkins University, 3400 N. Charles Street, Baltimore, MD 21218, USA}

\author[0000-0002-6562-8654]{Francesco Tombesi}
\affiliation{Physics Department, Tor Vergata University of Rome, Via della Ricerca Scientifica 1, 00133 Rome, Italy}
\affiliation{INAF --- Astronomical Observatory of Rome, Via Frascati 33, 00040 Monte Porzio Catone, Italy}
\affiliation{INFN --- Rome Tor Vergata, Via della Ricerca Scientifica 1, 00133 Rome, Italy}

\author[0000-0001-8496-4162]{Ruancun Li}
\affiliation{Kavli Institute for Astronomy \& Astrophysics and Department of Astronomy, Peking University, Beijing 100871, China}
\affil{Department of Astronomy, School of Physics, Peking University,
Beijing 100871, China}

\author[0000-0003-3703-5154]{Suvi Gezari}
\affiliation{Space Telescope Science Institute, 3700 San Martin Drive, Baltimore, MD 21218, USA}
\affiliation{Department of Physics and Astronomy, Johns Hopkins University, 3400 N. Charles Street, Baltimore, MD 21218, USA}

\author[0000-0003-3828-2448]{Javier~A.~Garc\'ia}
\affiliation{Astrophysics Science Division, NASA Goddard Space Flight Center, Greenbelt, MD 20771, USA}

\author[0000-0002-9589-5235]{Lixin Dai}
\affiliation{Department of Physics, University of Hong Kong, Pokfulam Road, Hong Kong}

\author[0000-0002-7706-5668]{Ryan Chornock}
\affiliation{Department of Astronomy, University of California, Berkeley, CA 94720, USA}

\author[0000-0002-1568-7461]{Wenbin Lu}
\affiliation{Department of Astronomy, University of California, Berkeley, CA 94720, USA}
\affiliation{Theoretical Astrophysics Center, University of California, Berkeley, CA 94720, USA}

\author[0000-0001-5390-8563]{S.~R.~Kulkarni}
\affiliation{Cahill Center for Astrophysics, California Institute of Technology, MC 249-17, 1200 E California Boulevard, Pasadena, CA 91125, USA}

\author[0000-0001-7115-2819]{Keith C. Gendreau}
\affiliation{Astrophysics Science Division, NASA Goddard Space Flight Center, Greenbelt, MD 20771, USA}

\author[0000-0003-1386-7861]{Dheeraj R. Pasham}
\affiliation{Kavli Institute for Astrophysics and Space Research, Massachusetts Institute of Technology, Cambridge, MA 02139, USA}

\author[0000-0003-1673-970X]{S. Bradley Cenko}
\affiliation{Astrophysics Science Division, NASA Goddard Space Flight Center, Greenbelt, MD 20771, USA}
\affiliation{Joint Space-Science Institute, University of Maryland, College Park, MD 20742, USA}

\author[0000-0003-0172-0854]{Erin Kara}
\affiliation{Department of Physics, Massachusetts Institute of Technology, Cambridge, MA 02139, USA}
\affiliation{Kavli Institute for Astrophysics and Space Research, Massachusetts Institute of Technology, Cambridge, MA 02139, USA}

\author[0000-0003-4768-7586]{Raffaella Margutti}
\affiliation{Department of Astronomy, University of California, Berkeley, CA 94720, USA}
\affiliation{Department of Physics, University of California, 366 Physics North MC 7300, Berkeley, CA 94720, USA}

\author[0009-0007-8764-9062]{Yukta Ajay}
\affiliation{Department of Physics and Astronomy, Johns Hopkins University, 3400 N. Charles Street, Baltimore, MD 21218, USA}

\author[0000-0002-4043-9400]{Thomas Wevers}
\affiliation{Space Telescope Science Institute, 3700 San Martin Drive, Baltimore, MD 21218, USA}

\author[0000-0003-0509-2541]{Tom M. Kwan}
\affiliation{Department of Physics, University of Hong Kong, Pokfulam Road, Hong Kong}


\author[0000-0002-8977-1498]{Igor Andreoni}
\affil{Joint Space-Science Institute, University of Maryland, College Park, MD 20742, USA}
\affil{Department of Astronomy, University of Maryland, College Park, MD 20742, USA}
\affil{Astrophysics Science Division, NASA Goddard Space Flight Center, Greenbelt, MD 20771, USA}

\author[0000-0002-7777-216X]{Joshua S. Bloom}
\affiliation{Department of Astronomy, University of California, Berkeley, CA 94720, USA}
\affiliation{Lawrence Berkeley National Laboratory, 1 Cyclotron Road, MS 50B-4206, Berkeley, CA 94720, USA}

\author[0000-0003-0228-6594]{Andrew J. Drake}
\affiliation{Cahill Center for Astrophysics, California Institute of Technology, MC 249-17, 1200 E California Boulevard, Pasadena, CA 91125, USA}

\author[0000-0002-3168-0139]{Matthew J. Graham}
\affiliation{Cahill Center for Astrophysics, California Institute of Technology, MC 249-17, 1200 E California Boulevard, Pasadena, CA 91125, USA}

\author[0000-0002-5698-8703]{Erica Hammerstein}
\affiliation{Department of Astronomy, University of Maryland, College Park, MD 20742, USA}

\author[0000-0003-2451-5482]{Russ R. Laher}
\affiliation{IPAC, California Institute of Technology, 1200 E. California Blvd, Pasadena, CA 91125, USA}

\author[0000-0002-2249-0595]{Natalie LeBaron}
\affiliation{Department of Astronomy, University of California, Berkeley, CA 94720, USA}

\author[0000-0003-2242-0244]{Ashish~A.~Mahabal}
\affiliation{Cahill Center for Astrophysics, California Institute of Technology, MC 249-17, 1200 E California Boulevard, Pasadena, CA 91125, USA}
\affiliation{Center for Data Driven Discovery, California Institute of Technology, Pasadena, CA 91125, USA}

\author[0000-0002-9700-0036]{Brendan O'Connor}
\affiliation{McWilliams Center for Cosmology and Astrophysics, Department of Physics, Carnegie Mellon University, Pittsburgh, PA 15213, USA}

\author[0000-0003-1227-3738]{Josiah Purdum}
\affiliation{Caltech Optical Observatories, California Institute of Technology, Pasadena, CA 91125, USA}

\author[0000-0002-7252-5485]{Vikram Ravi}
\affiliation{Cahill Center for Astrophysics, California Institute of Technology, MC 249-17, 1200 E California Boulevard, Pasadena, CA 91125, USA}

\author[0000-0001-8023-4912]{Huei Sears}
\affiliation{Center for Interdisciplinary Exploration and Research in Astrophysics (CIERA), Northwestern University, Evanston, IL 60202, USA}
\affiliation{Department of Physics and Astronomy, Northwestern University, Evanston, IL 60208, USA}

\author[0000-0003-4531-1745]{Yashvi Sharma}
\affiliation{Cahill Center for Astrophysics, California Institute of Technology, MC 249-17, 1200 E California Boulevard, Pasadena, CA 91125, USA}

\author[0000-0001-7062-9726]{Roger Smith}
\affiliation{Caltech Optical Observatories, California Institute of Technology, Pasadena, CA  91125}

\author[0000-0003-1546-6615]{Jesper Sollerman}
\affiliation{The Oskar Klein Centre, Department of Astronomy, Stockholm University, AlbaNova, SE-10691, Stockholm, Sweden}

\author[0000-0001-8426-5732]{Jean J.~Somalwar}
\affiliation{Cahill Center for Astrophysics, California Institute of Technology, MC 249-17, 1200 E California Boulevard, Pasadena, CA 91125, USA}

\author[0000-0002-9998-6732]{Avery Wold}
\affiliation{IPAC, California Institute of Technology, 1200 E. California Blvd, Pasadena, CA 91125, USA}

\begin{abstract}
We present the tidal disruption event (TDE) AT2022lri, hosted in a nearby ($\approx\!144$\,Mpc) quiescent galaxy with a low-mass massive black hole ($10^4\,M_\odot < M_{\rm BH} < 10^6\,M_\odot$). 
AT2022lri belongs to the TDE-H+He subtype.
More than 1\,Ms of X-ray data were collected with \nicer, \swift, and \xmm from 187\,d to 672\,d after peak. 
The X-ray luminosity gradually declined from $1.5\times 10^{44}\,{\rm erg\,s^{-1}}$ to $1.5\times 10^{43}\,{\rm erg\,s^{-1}}$ and remains much above the UV and optical luminosity, consistent with a super-Eddington accretion flow viewed face-on.
Sporadic strong X-ray dips atop a long-term decline are observed, with variability timescale of $\approx\!0.5$\,hr--1\,d and amplitude of $\approx\!2$--8. 
When fitted with simple continuum models, the X-ray spectrum is dominated by a thermal disk component with inner temperature going from $\sim\! 146$\,eV to $\sim\! 86$\,eV. 
However, there are residual features that peak around 1\,keV, which, in some cases, cannot be reproduced by a single broad emission line. 
We analyzed a subset of time-resolved spectra with two physically motivated models describing either a scenario where ionized absorbers contribute extra absorption and emission lines or where disk reflection plays an important role.
Both models provide good and statistically comparable fits, show that the X-ray dips are correlated with drops in the inner disk temperature, and require the existence of sub-relativistic (0.1--0.3$c$) ionized outflows.
We propose that the disk temperature fluctuation stems from episodic drops of the mass accretion rate triggered by magnetic instabilities or/and wobbling of the inner accretion disk along the black hole's spin axis.
\end{abstract}

\vspace{1em}

\section{Introduction} 
\subsection{Super-Eddington Accretion onto Low-mass Massive Black Holes in Tidal Disruption Events}\label{subsec:intro_superEdd}

The detection of $\sim\!10^9\,M_\odot$ accreting black holes (BHs) at merely $<\!1$\,Gyr after the big bang \citep{Fan2023} implies that a sustained period of super-Eddington accretion must have occurred in the early Universe. However, the process of BH super-Eddington accretion is observationally poorly characterized due to the fact that the majority of active galactic nuclei (AGN) are accreting at sub-Eddington rates. 

Stars that are tidally disrupted outside the event horizon of a massive black hole (MBH) give rise to a panchromatic transient as the disrupted material falls onto the MBH (see \citealt{Gezari2021} for a recent review). 
Basic theory for such tidal disruption events (TDEs)  predicts that after the disruption of a star (with mass $M_\ast$ and radius $R_\ast$), the mass fall-back rate $\dot M_{\rm fb}$ initially rises for a fall-back timescale of 
\begin{align}
    t_{\rm fb} = 41\,{\rm d} (M_{\rm BH}/10^6\,M_\odot)^{1/2} m_\ast^{-1} r_\ast^{3/2}
\end{align}
to a peak mass fall-back rate of
\begin{align}
    \frac{\dot M_{\rm fb, peak}}{\dot M_{\rm Edd}} & = 136 \eta_{-1}m_\ast^2 r_\ast^{-3/2} (M_{\rm BH}/10^6\,M_\odot)^{-3/2},\\
    \dot M_{\rm Edd} & = \frac{L_{\rm Edd}}{\eta c^2},\;\; L_{\rm Edd} = 1.26\times 10^{38} (M_{\rm BH}/M_\odot)\,{\rm erg\,s^{-1}}
\end{align}
where $m_\ast = M_\ast/M_\odot$, $r_\ast = R_\ast / R_\odot$, $\eta$ is the accretion radiative efficiency, and $\eta_{-1} \equiv \eta/0.1$.
The returned stellar debris quickly circularize to form a compact accretion flow on the timescale of $t_{\rm circ}\lesssim\! t_{\rm fb}$ \citep{Bonnerot2016, Hayasaki2016, Bonnerot2020, Andalman2022, Steinberg2024}. For lower-mass MBHs ($M_{\rm BH}\lesssim 10^6\,M_\odot$), $\dot M_{\rm fb}$ remains super-Eddington for a few years --- radiation is trapped in the disk and puffs up the disk, such that the disk thickness over radius is on the order of unity \citep{Abramowicz1988, Dai2021}. 
In this case, the viscous timescale is short and the mass accretion rate is expected to closely follow the $\dot M_{\rm fb}\propto t^{-5/3}$ evolution that samples a wide range in a known way \citep{Rees1988, Phinney1989, Evans1989, Lodato2009, DeColle2012}. 
It is for this reason that TDEs (especially those occurring in lower-mass BHs) provide clean laboratories to understand BH super-Eddington accretion.

While the early-time TDE emission might be powered either by reprocessed disk emission at larger radii \citep{Loeb1997, Roth2016, Metzger2016, Dai2018, Metzger2022} or adiabatic photon trapping in expanding outflows \citep{Jiang2016_self_crossing_shock, Lu2020}, it is generally expected that direct disk emission should become the dominant component after a ${\rm few}\times t_{\rm fb}$.
A rough estimate for the inner disk temperature is the temperature associated with an Eddington luminosity from an order of unity factor times the gravitational radius \citep{Ulmer1999}:
\begin{align}
    T_{\rm in} \approx \left[\frac{L_{\rm Edd}}{4\pi (6R_g)^2 \sigma_{\rm SB}} \right]^{1/4} 
    = 59\,{\rm eV} (M_{\rm BH} / 10^6\,M_\odot)^{-1/4}.
\end{align}
In lower-mass MBHs, the disk emission should peak in the soft X-ray band. 
Observationally, however, the soft X-ray emission of many optically-selected TDEs in lower-mass MBHs remain much fainter than the UV and optical emission out to $t\gtrsim 1$\,yr\footnote{For example, see AT2018lna, AT2018zr, AT2020vwl, AT2020wey \citep{Guolo2024}, AT2020neh \citep{Angus2022}, and AT2020vdq \citep{Somalwar2023_20vdq}.}, indicating that the TDE structure is more complex than predicted by basic theory. Possible explanations for the lack of X-ray emission include a viewing-angle dependence of the optical-to-X-ray ratio for a thick disk \citep{Dai2018, Metzger2022}, insufficient debris circularization at early times \citep{Shiokawa2015}, and significantly reduced mass inflow rate at the inner disk region due to mass loss in line-driven disk winds \citep{Miller2015_disk_wind}.

\subsection{AT2022lri}
AT2022lri ($\alpha = 02^{\rm h}20^{\rm m}8^{\rm s}.01$, $\delta = -22^{\circ}42^{\prime}15^{\prime\prime}.21$) was first reported to the Transient Name Server (TNS) by the ATLAS team (as ATLAS22pnz) in June 2022 with an $o$-band detection at 17.2\,mag \citep{Tonry2022_discovery_report}. It was subsequently detected by the Zwicky Transient Facility (ZTF; \citealt{Bellm2019b, Graham2019, Dekany2020}) survey (as ZTF22abajudi) on 2022 August 14 (MJD 59805.508) with an $r$-band detection at $18.77\pm0.04$\,mag. 
On 2022 September 3, this object passed the ZTF Bright Transient Survey (BTS; \citealt{Fremling2020, Perley2020}) experiment. On 2022 September 17, it passed our TDE selection filter\footnote{Before September 2023, the ZTF team selected nuclear transients by filtering public alerts with the \texttt{AMPEL} broker \citep{Nordin2019}.} \citep{vanVelzen2019} and optical spectroscopy with the Double Spectrograph (DBSP; \citealt{Oke1982}) was triggered on 2022 October 11.
An initial Neil Gehrels Swift Observatory (hereafter \swift) snapshot was triggered on 2022 October 13, which revealed a bright UV source as well as bright and soft X-ray emission detected by the X-Ray Telescope (XRT; \citealt{Burrows2005}).
On 2022 October 24, we classified AT2022lri as a TDE \citep{Yao2022_AN_22lri} based on characteristic broad emission lines shown in its DBSP spectrum \citep{Yao2022_CR}, the blue color ($uvw2-r\approx\!-1$\,mag), and the luminous soft X-ray detection.

From 2022 October 23 to 2024 March 5, we conducted high-cadence X-ray observations using the \textit{Neutron Star Interior Composition Explorer} (\nicer; \citealt{Gendreau2016}) (for a total of 845\,ks), the XRT (87\,ks), and three \xmm exposures (152\,ks).
Hosted by a low-mass MBH, the X-ray light of AT2022lri remains much in excess of the UV and optical emission. The X-ray luminosity alone is $\gtrsim L_{\rm Edd}$, in line with the basic TDE theory (\S\ref{subsec:intro_superEdd}). 
What distinguishes AT2022lri from all previously known TDEs is the following:
\begin{itemize}
    \item The observed peak 0.3--2\,keV flux reached $4.7\times 10^{-11}\,{\rm erg\,s^{-1}\,cm^{-2}}$ (1.6\,mCrab), making it brighter than all other non-jetted TDEs in the literature. The bright X-ray flux facilitated high-quality X-ray spectroscopy, showing evidence of sub-relativistic outflows.
    \item The intensive X-ray monitoring campaign, particularly with \nicer, uncovered fast large-amplitude (multiplicative factors of $\sim\!2$--8) X-ray variability on hour--day timescales. The X-ray spectral evolution within such short timescales exhibit rapid inner disk temperature variations. 
\end{itemize}
These distinct X-ray characteristics of AT2022lri provide novel insights into the inner accretion flow of TDEs under super-Eddington accretion.

This paper is structured as follows. 
\S\ref{sec:hostgalaxy} presents the observation and analysis of AT2022lri's host galaxy. 
In \S\ref{sec:uvopt}, we outline the UV and optical observations, data reduction, and analysis of AT2022lri. 
In \S\ref{sec:xray}, we provide a detailed description of the X-ray observations, data processing, and spectral modeling. 
In \S\ref{sec:discuss}, we discuss our results in context of BH super-Eddington accretion, and compare AT2022lri with other nuclear transients with similar properties. 
Finally, we summarize our results in \S\ref{sec:summary}.

UT time is used throughout the paper. 
We adopt a standard $\Lambda$CDM cosmology with matter density $\Omega_{\rm M} = 0.3$, dark energy density $\Omega_{\Lambda}=0.7$, and the Hubble constant $H_0=70\,{\rm km\,s^{-1}\,Mpc^{-1}}$, implying a luminosity distance to AT2022lri of $D_L = 143.8\,{\rm Mpc}$ at the redshift of $z=0.03275\pm0.00001$ (see \S\ref{subsec:velocity_dispersion}). 
UV and optical magnitudes are reported in the AB system. 
We use the extinction law from \citet{Cardelli1989}, assume $R_V = 3.1$, and adopt a Galactic extinction of $E_{B-V, {\rm MW}} =  0.0158$\,mag \citep{Schlafly2011}. 
Uncertainties are reported at the 68\% confidence intervals, and upper limits are reported at $3\sigma$. 
Coordinates are given in J2000.

\section{Host Galaxy Observations and Analysis}
\label{sec:hostgalaxy}

\subsection{Velocity Dispersion} \label{subsec:velocity_dispersion}

A medium-resolution spectrum was obtained on 2023 August 26.6 
using the Echellette Spectrograph and Imager (ESI; \citealt{Sheinis2002}) on the Keck II telescope. 
At this phase, the TDE flux is negligible in the optical band compared with the host galaxy flux (see \S\ref{sec:uvopt}). The observation was performed in the Echellette mode with a 0.75$^{\prime\prime}$ slit, which gives a resolving power of $R = 5350$ (i.e., $\sigma_{\rm inst}=24\,{\rm km\,s^{-1}}$). The data was processed using the \texttt{makee} pipeline following standard procedures. From the median light profile of the traces, we estimated that the half-light radius ($r_{1/2}$) of the galaxy is $\approx 0.98^{\prime\prime}$. We then extracted the ESI spectrum using a radius of $r_{1/2}$ (i.e., 6.2\,pixels), and normalized the spectrum by fitting a spline function to the continuum, with prominent emission and absorption lines masked. 

\begin{figure*}[htbp!]
    \centering
    \includegraphics[width=0.9\textwidth]{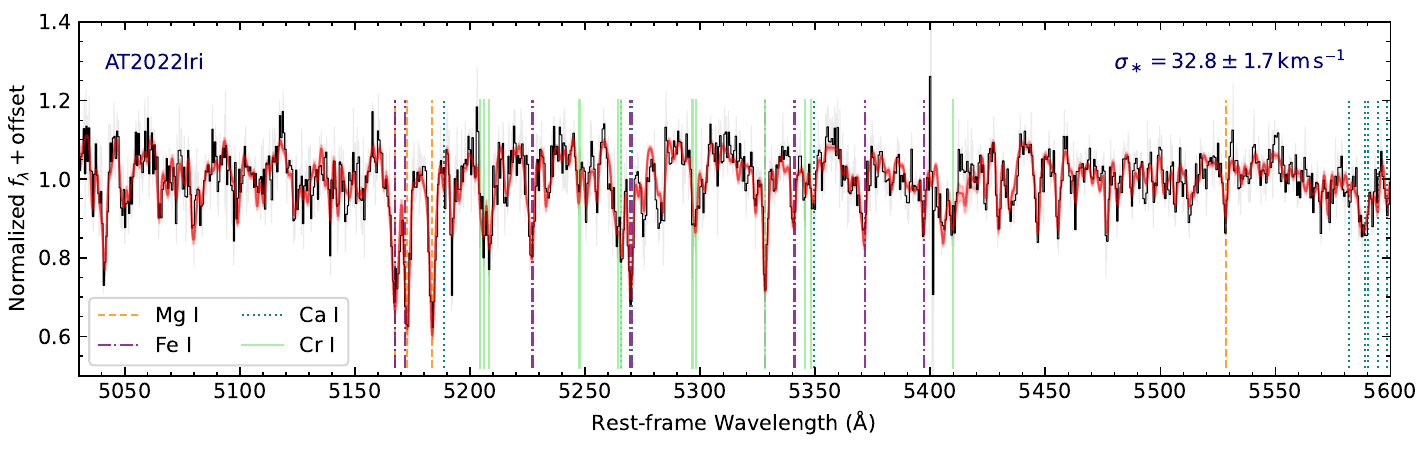}
    \caption{ESI spectrum of AT2022lri's host galaxy (black) and the best-fit model (red). Prominent host galaxy absorption lines are indicated by the vertical lines. The median S/N of the fitted wavelength range of 5030--5600\,\AA\ is 15.7. \label{fig:ESI_spec}}
\end{figure*}

To infer the stellar velocity dispersion ($\sigma_\ast$), we modeled the ESI spectrum with penalized pixel-fitting (\texttt{pPXF}; \citealt{Cappellari2004, Cappellari2017}) and the ELODIE v3.1 high-resolution ($R=42000$) library \citep{Prugniel2001, Prugniel2007}, following the same steps adopted by \citet{Yao2023}. 
Fitting over the wavelength range of 5030--5600\,\AA, we derived a redshift of $z=0.03275\pm0.00001$ and a velocity dispersion of $\sigma_{\ast} = 32.8\pm1.7\,{\rm km\,s^{-1}}$. 
The fitting result is shown in Figure~\ref{fig:ESI_spec}. 
If using a wider wavelength range of 4600--5600\,\AA, the velocity dispersion remains consistent at $\sigma_{\ast} = 32.0\pm1.6\,{\rm km\,s^{-1}}$.

\subsection{Morphology and Spectral Energy Distribution Fitting} \label{subsec:galfits}

\begin{figure}[htbp!]
\centering
\includegraphics[width=0.9\textwidth]{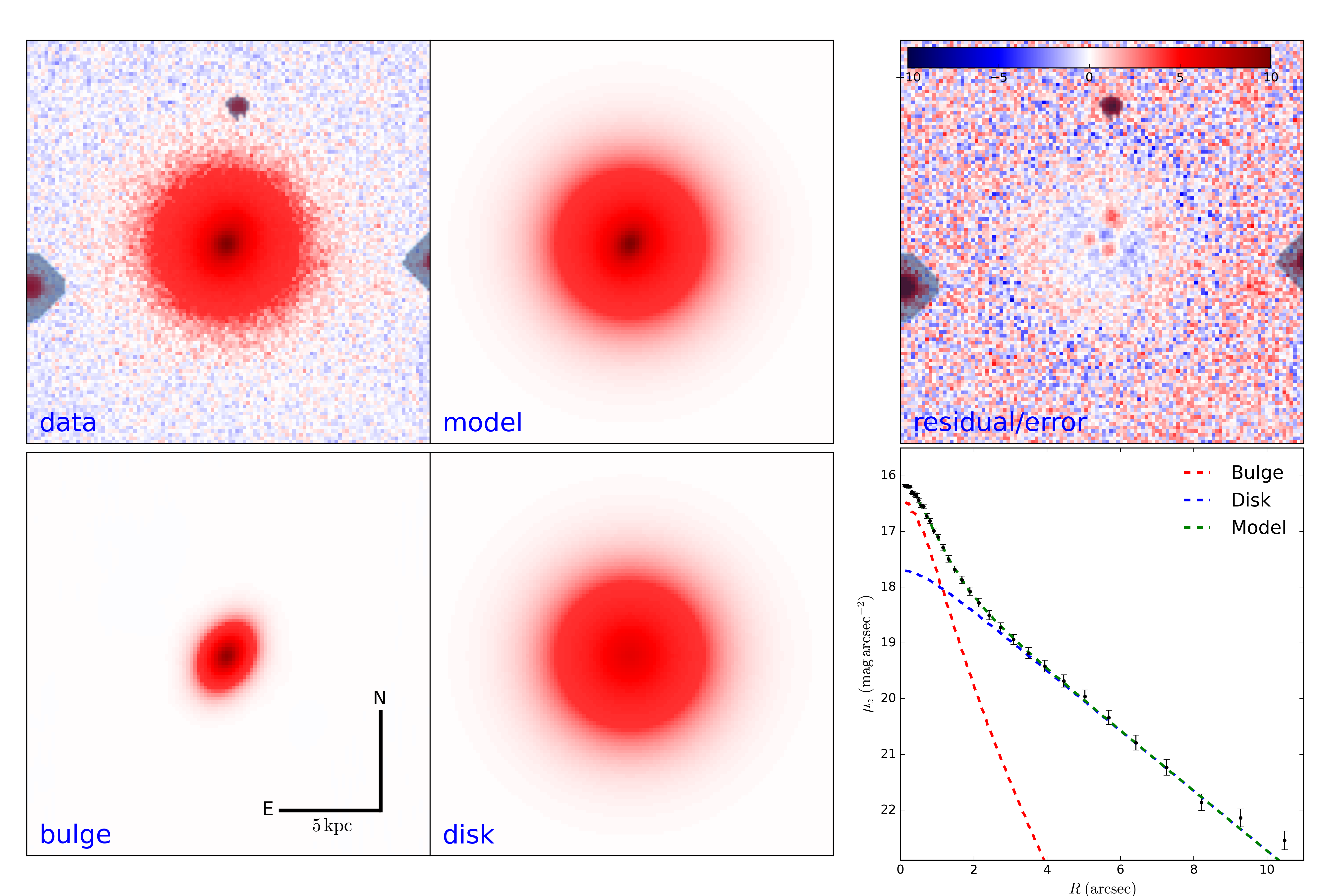}
\caption{Results of the \textsc{GALFITS} imaging decomposition for the LS $z$-band image of AT2022lri's host galaxy. 
 {\it Upper left}: The LS $z$-band image centered on AT2022lri, with a size of 30\arcsec. The shaded region represents a mask for contaminating sources. 
 {\it Upper middle}: The model image generated using the best-fit \textsc{GALFITS} parameters detailed in Table~\ref{tab:host}. 
 {\it Upper right}: The error-normalized residual image, obtained after subtracting the best-fit \textsc{GALFITS} model. 
 {\it Lower left}: The model image of the bulge component.
 {\it Lower middle}: The model image of the disk component. 
 {\it Lower right}: The surface brightness profile of the observed $z$-band image (black error bars), the bulge model (red), the disk model (blue), and the total \textsc{GALFITS} model (green).  
		\label{fig:host_image}}
\end{figure}

We employed the multi-band imaging decomposition tool \textsc{GALFITS} (Li \& Ho, in prep) to analyze the pre-TDE images of AT2022lri. The dataset comprises UV images from the Galaxy Evolution Explorer (GALEX, \citealp{Martin2005}), $griy$ images from Pan-STARRS1 \citep{Chambers2016}\footnote{The Pan-STARRS1 $z$-band image was excluded due to artifacts resulting from overexposure.}, $grz$ images from the DESI Legacy Imaging Survey (LS; \citealt{Dey2019}), as well as infrared images from the Two-Micron All Sky Survey \citep{Skrutskie2006} and the Wide-field Infrared Survey Explorer \citep{Wright2010}, totaling 16 images. 
\textsc{GALFITS} performs morphological decomposition and spectral energy distribution (SED) fitting simultaneously by assigning different components distinct structural parameters as well as stellar populations. For GALEX images, we collected the entire image tile encompassing AT2022lri. For other images, a cutout of 4000\arcsec\ was utilized. This approach ensures adequate sampling of sky pixels, facilitating accurate sky subtraction.
For each image, 
we generated a point spread function (PSF) by fitting stars within the same field. These stars were selected by cross-matching with the Gaia DR3 catalog \citep{Gaiadr32023AA}. Subsequently, a 30\arcsec\ box cutout was applied to all 16 images to serve as the input data for \textsc{GALFITS}. Additionally, we generated a common mask image for all bands to minimize the impact of contaminating sources on \textsc{GALFITS}. 
The upper left panel of Figure~\ref{fig:host_image} shows the LS $z$-band image, where the mask image is depicted as the shaded region.  

During the \textsc{GALFITS} fitting, we tested models featuring both single and double \sersic\ components. To define each component, besides two parameters for the position, there are four morphological parameters and five SED-related parameters. The morphological parameters are the \sersic\ index ($n$), effective radius ($R_\mathrm{e}$), axis ratio ($q$), and position angle (PA), where PA is measured counterclockwise from the North direction. For the SED, we assume the stellar populations consist of two parts: one single stellar population (a burst) and another with a constant star formation rate since $z\simeq 15$. The SED's free parameters are the stellar mass ($M_\ast$) of the component, the age of the first stellar population ($t_0$), the star formation rate (SFR) of the second stellar population, attenuation from dust in the $V$ band ($A_V$), and stellar metallicity ($Z$). \textsc{GALFITS} also considers the contribution from nebular emission to the SED. We fixed the gas phase metallicity to be the same as the stellar metallicity and set the ionization parameter $U = 10^{-3}$, which are tested to have negligible effects on the fitting. 
Nested sampling is used for posterior distribution estimation, where all parameters are assumed to have a uniform distribution, utilizing the \textsc{PYTHON} package \texttt{dynesty}. 
Based on our tests, the two-component model demonstrates a reduced chi-squared value of $\chi_{\rm r}^2 \simeq 0.33$, significantly better than that of the single-component model ($\chi_{\rm r}^2 \simeq 2.09$). The best-fit parameters from analysis of 16 broadband images
are shown in Table~\ref{tab:host}.
The \textsc{GALFITS} model for the LS $z$-band image is shown as an example in Figure~\ref{fig:host_image}.

\begin{deluxetable}{cccccccccc}[htbp!]
\tablecaption{Properties of the host galaxy} \label{tab:host}
\tabletypesize{\footnotesize}
\tablehead{
      \colhead{Component}             &
	\colhead{$\log{M_\ast}$}   &
    \colhead{$n$} &
    \colhead{$R_\mathrm{e}$} &
    \colhead{$q$} &
    \colhead{PA} &
      \colhead{SFR} &
      \colhead{$t_0$} &
      \colhead{$A_V$} &
      \colhead{$Z$} \\
&
\colhead{($M_\odot$)} &
    &
\colhead{(\arcsec) } &
    &
 \colhead{($^{\circ}$)} &   
\colhead{($M_\odot\, \rm yr^{-1}$)} &
\colhead{(Gyr)} &
\colhead{(mag)}    &
\colhead{($Z_\odot$)} 
}
\startdata
      bulge & $8.97\pm0.04$ & $1.93\pm0.11$ & $0.87\pm0.03$ & $0.62\pm0.02$ & $-28.9\pm1.1$ & $< 3\times 10^{-5}$ &$1.08\pm 0.02$ & $0.34\pm 0.06$ & $0.40\pm0.12$\\
      disk & $9.46\pm0.03$ & $0.92\pm0.04$ & $3.63\pm0.05$ & $0.96\pm0.01$ & $-89.4\pm1.3$ & $< 1\times 10^{-4}$ &$1.41\pm 0.02$ & $0.31\pm 0.03$ & $0.20\pm0.06$\\
     \enddata
\end{deluxetable}

In Table~\ref{tab:host}, the first component has a larger \sersic\ index ($n\simeq 1.93$) but a smaller size ($R_\mathrm{e} \sim 0.9\arcsec$), while the second component has a \sersic\ index of approximately $0.92$ but a larger size ($R_\mathrm{e} \sim 3.6\arcsec$). Consequently, we identify the first component as the stellar bulge and the second as the stellar disk. These components also exhibit notably different axis ratios; the disk component has $q\simeq 0.96$, suggesting a face-on view. Given that classical bulges typically have higher \sersic\ indices than pseudo bulges \citep{Kormendy2004ARAA,Gao2020ApJS}, a bulge with $n\lesssim 2$ is usually indicative of a pseudo bulge \citep{Fisher2008AJ}. Additionally, pseudo bulges are often less spherical compared to classical bulges \citep{Gao2020ApJS}. Therefore, the bulge of AT2022lri, with $n\simeq 1.93$ and $q\simeq 0.62$, is more likely to be a pseudo bulge. 

\subsection{Black Hole Mass} \label{subsec:Mbh}
We infer the black hole mass $M_{\rm BH}$ using host galaxy scaling relations. 
Notably, the velocity dispersion of $\sigma_{\ast} = 32.8\pm1.7\,{\rm km\,s^{-1}}$ (\S\ref{subsec:velocity_dispersion}) measured for AT2022lri's host galaxy is lower than any of the 29 TDE hosts' $\sigma_\ast$ values summarized in the review article of \citet{French2020}, and the 19 $\sigma_\ast$ measurements from the more recent ZTF sample by \citet{Yao2023}.
To our knowledge, no other TDE host has been recorded with a lower measured velocity dispersion.
Using the \citet{Kormendy2013} $M_{\rm BH}$--$\sigma_\ast$ relation, we estimate ${\rm log}(M_{\rm BH}/M_\odot) = 5.05\pm0.25({\rm stat})\pm0.29({\rm sys})$. Using the \citet{Greene2020} $M_{\rm BH}$--$\sigma_\ast$ relation for late-type galaxies, we obtain ${\rm log}(M_{\rm BH}/M_\odot) = 4.61\pm0.42({\rm stat})\pm0.58({\rm sys})$. Using the \citet{Ferrarese2005} $M_{\rm BH}$--$\sigma_\ast$ relation, we have ${\rm log}(M_{\rm BH}/M_\odot) = 4.40\pm0.36$.

The total stellar mass of the host is ${\rm log}(M_{\rm gal}/M_\odot)=9.58\pm0.03$ (sum of the bulge and disk masses in Table~\ref{tab:host}). 
This is lower than 78\% (29/37) of the TDE hosts' $M_{\rm gal}$ values in \citet{French2020}, and lower than 85\% (28/33) of $M_{\rm gal}$ measurements from \citet{Yao2023}. 
Using the \citet{Greene2020} $M_{\rm BH}$--$M_{\rm gal}$ relation for late-type galaxies, we obtain ${\rm log}(M_{\rm BH}/M_\odot) = 5.26\pm0.25({\rm stat})\pm0.65({\rm sys})$.
Separately, using the bulge mass of $M_{\rm bulge}=10^{8.97\pm0.04}\,M_\odot$ (\S\ref{subsec:galfits}) and the $M_{\rm BH}$--$M_{\rm bulge}$ correlation for pseudo bulges \citep{Kormendy2013, Li2022_1ES}, we estimate ${\rm log}(M_{\rm BH}/M_\odot) = 5.88\pm0.05({\rm stat})\pm0.33({\rm sys})$. 

To summarize, the $M_{\rm BH}$ values estimated from $\sigma_{\ast}$ is generally smaller\footnote{This difference is possibly caused by a velocity dispersion drop in the central region of the host, which is not uncommon in S0 galaxies or galaxies in the mass bin of ${9.5}<{\rm log}(M_{\rm gal}/M_\odot)<10.0$ \citep{Ouellette2022}.} than that estimated using $M_{\rm gal}$ and $M_{\rm bulge}$. 
It's important to note that these estimations are influenced by inherent uncertainties in the scaling relations, which may stem from calibration uncertainty and increased dispersions at the low-mass end.
In the following, we adopt an intermediate value of ${\rm log}(M_{\rm BH}/M_\odot) \approx 5$. while acknowledging that the uncertainty associated with this $M_{\rm BH}$ estimate can be as large as 1\,dex.

\subsection{Historical X-ray Constraints}

We obtained constraints on the historical X-ray luminosity of AT2022lri's host galaxy. The host was not detected by the \rosat all sky survey \citep{Truemper1982, Voges1999} in 1990/1991 with a 3$\sigma$ 0.1--2.4\,keV upper limit of $<0.303\,{\rm count\,s^{-1}}$. Assuming an absorbed powerlaw spectral shape with Galactic $N_{\rm H}$ and $\Gamma=2$, we obtain a 0.3--2\,keV flux upper limit of $<2.3\times 10^{-12}\,{\rm erg\,s^{-1}\,cm^{-2}}$. Assuming an absorbed blackbody spectral shape with Galactic $N_{\rm H}$ and $kT_{\rm bb}=0.1$\,keV gives a similar upper limit of $<2.1\times 10^{-12}\,{\rm erg\,s^{-1}\,cm^{-2}}$.

On 2008 January 16, the field of AT2022lri was observed by \swift/XRT. We estimated a 0.3--10\,keV $3\sigma$ upper limit of $4.80\times 10^{-3}$\,\cps. This corresponds to a 0.3--2\,keV flux limit of $<7.0\times10^{-14}\,{\rm erg\,s^{-1}\,cm^{-2}}$ and $<1.1\times10^{-13}\,{\rm erg\,s^{-1}\,cm^{-2}}$ for a power-law and blackbody spectral shape, respectively. 

To summarize, the host galaxy is X-ray faint, with a flux upper limit of $\lesssim 10^{-13}\,{\rm erg\,s^{-1}\,cm^{-2}}$, which corresponds to a luminosity upper limit of $\lesssim 2.5\times 10^{41}\,{\rm erg\,s^{-1}}$. This further suggests that the host of AT2022lri does not contain an AGN, consistent with the lack of diagnostic AGN forbidden lines (such as [\ion{O}{III}] and [\ion{N}{II}]) in the very late-time optical spectra (see 
\S\ref{subsec:opt_spec}).

\section{UV--Optical Observations and Analysis} \label{sec:uvopt}

\subsection{Optical Photometry}\label{subsec:opt_phot}

\begin{figure*}[htbp!]
    \centering
    \includegraphics[width=0.95\textwidth]{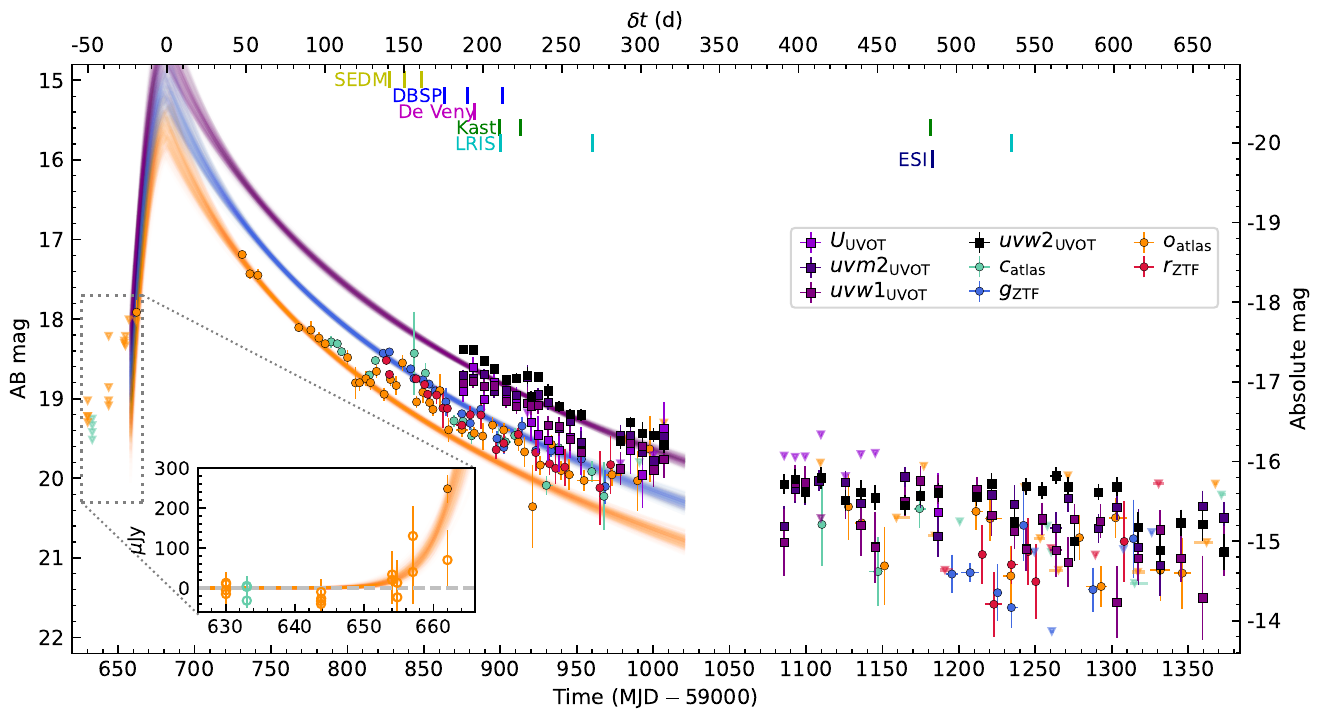}
    \caption{UV and optical light curves of AT2022lri (data points are $>1.5\sigma$ measurements, semitransparent downward triangles are $3\sigma$ upper limits). 
    The solid lines are single-temperature Gausssian-rise exponential-decline models in three bands ($o_{\rm ATLAS}$, $g_{\rm ZTF}$, and $uvw1_{\rm UVOT}$) fitted to the first-year data. 
    The ticks along the upper axis marks the epoch of optical spectroscopy (\S\ref{subsec:opt_spec}). 
    The inset shows the ATLAS light curve zoomed around time of first detection.
    \label{fig:lc_uvopt}}
\end{figure*}

We obtained ZTF and ATLAS forced photometry \citep{Masci2019, Tonry2018, Smith2020, Shingles2021} using the median position of all ZTF alerts up to MJD\,59940. 
Baseline correction was performed following the procedures outlined in \citet{Yao2019}.
The Galactic extinction-corrected optical light curves are shown in Figure~\ref{fig:lc_uvopt}, and presented in Appendix~\ref{sec:table_log} (Table~\ref{tab:phot}).

The forced ATLAS photometry reveals a 7$\sigma$ detection on 2022 March 24 (MJD 59662.006) at $o=17.91\pm0.15$\,mag that precedes the ATLAS TNS report\footnote{We note that there were no other similarly significant detections in the pre-outburst ATLAS historical light curve.}. This marks the first optical detection and the start of the TDE rise (see the inset of Figure~\ref{fig:lc_uvopt}).
However, the peak of the optical light curve was missed due to Sun occultation. 
Fitting the first-year light curve with a Gaussian rise and a power-law decline yields a peak time around MJD 59682.
Hereafter we use $\delta t$ to denote rest-frame days relative to MJD 59682.
We binned the optical light curves at $\delta t>0$ by 1--14\,days to reveal sub-threshold optical detections at late time, and to facilitate SED modeling with the ultraviolet data (see details in \S\ref{subsec:bbfit}). 

\subsection{Swift/UVOT Photometry}  \label{subsec:uvot}
AT2022lri was observed by the Ultra-Violet/Optical Telescope (UVOT; \citealt{Roming2005}) on board \swift under a series of time-of-opportunity (ToO) requests submitted by Y. Yao. 

UVOT observations were obtained with the ``$U$+All UV'' filters or the ``All UV'' filters.
Some observations were split into two obsIDs. We stacked these adjacent obsIDs with \texttt{uvotimsum} to improve the sensitivity.
The source flux was measured with \texttt{uvotsource}, using a circular region with $r_{\rm src} = 9^{\prime\prime}$. The relatively large radius was chosen to ensure the capture of all the host galaxy flux.
The background flux was measured using four nearby circular source-free regions with $r_{\rm bkg}=10^{\prime\prime}$. 
We estimated host galaxy flux within a $9^{\prime\prime}$ aperture using a host galaxy synthesis model, which was constructed following the same procedures adopted by \citet{vanVelzen2021} and \citet{Hammerstein2023}.
Figure~\ref{fig:lc_uvopt} shows the host subtracted UVOT photometry. 
Our final UVOT photometry is presented in Appendix~\ref{sec:table_log} (Table~\ref{tab:phot}).

\subsection{Modeling the UV and Optical Photometry} \label{subsec:bbfit}

We modeled the UV and optical photometry of AT2022lri with a blackbody function following the method adopted by \citet{Yao2020}. We fitted for both the blackbody radius ($R_{\rm bb}$) and the blackbody temperature ($T_{\rm bb}$) for epochs where two conditions are met: (1) the number of filters with significant detections (signal-to-noise ratio, ${\rm SNR}>3$) is greater than three, and (2) there is at least one significant detection in the optical (ZTF or ATLAS) band. A total of 20 epochs met these conditions. 
For the remaining 23 epochs that do not satisfy the above conditions, we only fitted for $R_{\rm bb}$, while fixing the $T_{\rm bb}$ value to that of the closest epoch with fitted $T_{\rm bb}$ estimate. 

Generally speaking, the best-fit $T_{\rm bb}$ remains around $2.5\times 10^4$\,K, which is typical for ZTF-selected TDEs \citep{Hammerstein2023, Yao2023}. The blackbody luminosity $L_{\rm bb}$ decreased from $\approx 10^{43}\,{\rm erg\,s^{-1}}$ to $\approx 10^{42}\,{\rm erg\,s^{-1}}$, and the blackbody radius $R_{\rm bb}$ decreased from $\approx\!2\times10^{14}$\,cm to $\approx\!6\times10^{13}$\,cm. These measurements are typical for optically-selected TDEs \citep{vanVelzen2020, Hammerstein2023}. 

\begin{figure*}[htbp!]
    \centering
    \includegraphics[width=0.54\textwidth]{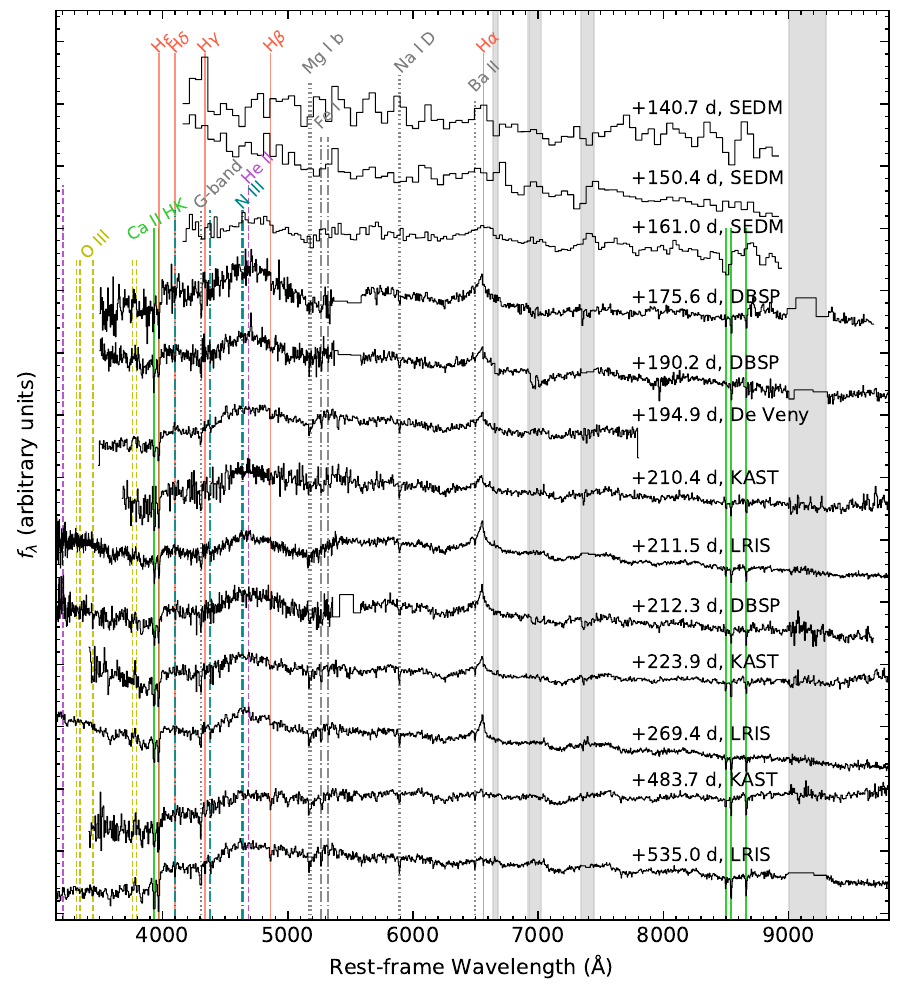}
    \includegraphics[width=0.34\textwidth]{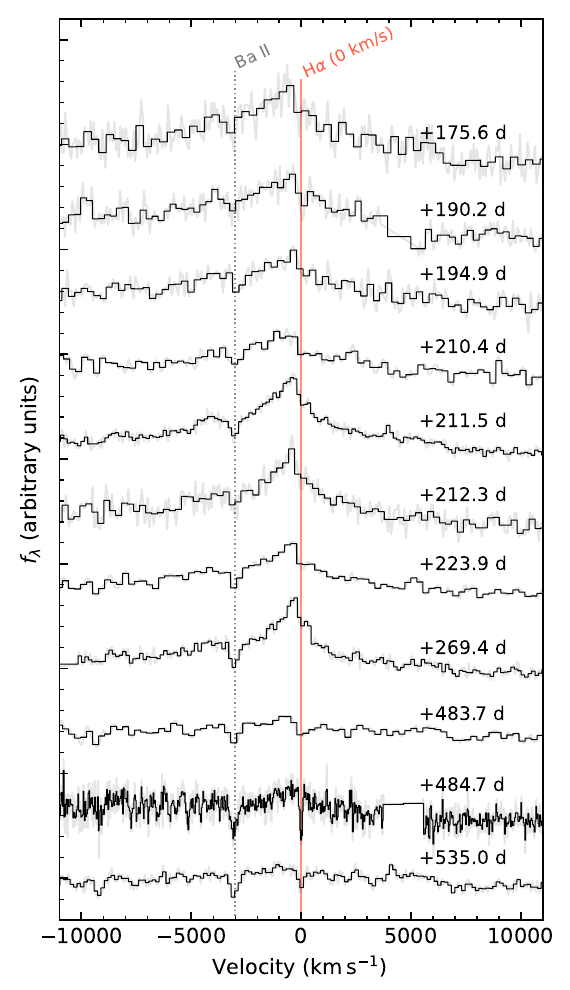}\\
    \includegraphics[width=0.54\textwidth]{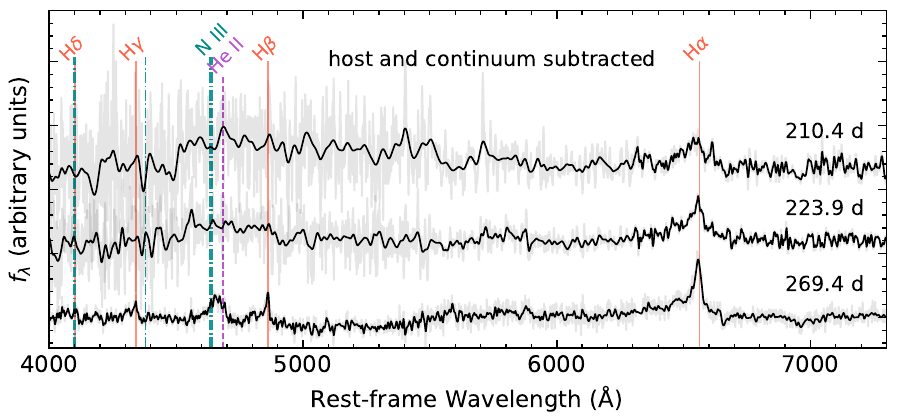}
    \includegraphics[width=0.34\textwidth]{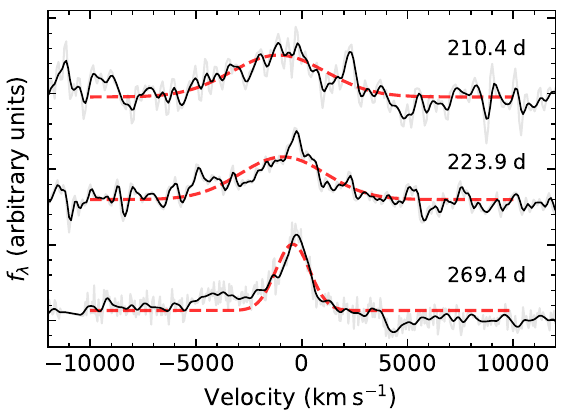}
    \caption{
    \textit{Upper left}: Observed low-resolution optical spectra of AT2022lri. 
    The vertical lines mark host galaxy lines and spectral lines common in TDEs. The gray bands mark atmospheric telluric features. Strong telluric features have been masked. 
    \textit{Upper right}: Zoom-in to the H$\alpha$ region of the observed spectra in velocity space. The ESI spectrum ($\delta t = 484.7$\,d, \S\ref{subsec:velocity_dispersion}) is also shown for comparison. 
    \textit{Lower left}: The translucent lines show host and continuum subtracted spectra, with the overlying black lines showing the same spectra convolved with Gaussian kernels for visual clarity. For the LRIS spectrum at +269.4\,d, we use a kernel FWHM of $200\,{\rm km\,s^{-1}}$. Since the Kast spectra are of lower SNR, we use FWHM kernel sizes of $1500\,{\rm km\,s^{-1}}$ ($1000\,{\rm km\,s^{-1}}$) on the blue side at $\lambda_{\rm rest}\lesssim 6000$\,\AA, and $300\,{\rm km\,s^{-1}}$ ($200\,{\rm km\,s^{-1}}$) on the red side for the +210.4\,d (+223.9\,d) data.
    \textit{Lower right}: Zoom-in to the H$\alpha$ region of the host and continuum subtracted spectra in velocity space. The dashed red lines are Gaussian fits to the line profile.  
    \label{fig:opt_spec}}
\end{figure*}

\subsection{Low-resolution Optical Spectroscopy} \label{subsec:opt_spec}

Optical spectroscopic observations were coordinated in part with the \texttt{Fritz} platform \citep{Coughlin2023} developed upon the \texttt{SkyPortal} software \citep{vanderWalt2019}.
Epochs of optical spectroscopy are marked along the upper axis of Figure~\ref{fig:lc_uvopt}. 

We obtained low-resolution optical spectroscopy with 
the Spectral Energy Distribution Machine (SEDM, \citealt{Blagorodnova2018}, \citealt{Rigault2019}, \citealt{Kim2022}) on the robotic Palomar 60 inch telescope (P60, \citealt{Cenko2006}),
the DBSP on the Palomar 200-inch Hale telescope (P200), 
the De Veny Spectrograph on the Lowell Discovery Telescope (LDT),
the Kast spectrograph on the Shane 3-m telescope at Lick Observatory \citep{millerstone93},
and the Low Resolution Imaging Spectrograph (LRIS; \citealt{Oke1995}) on the Keck-I telescope.
A log is given in Appendix~\ref{sec:table_log} (Table~\ref{tab:spec}).
Instrumental setup and data reduction of DBSP, De Veny, and LRIS spectra are the same as outlined in Appendix~B of \citet{Yao2022_21ehb}. 
The Kast spectra used the D57 dichroic with the 600/4310 grism on the blue arm and the 300/7500 grating on the red side. The Kast reduction was performed in a similar manner to that outlined by \citet{silverman2012}.
All low-resolution spectra are shown in the upper left panel of Figure~\ref{fig:opt_spec}. 

To reveal weak TDE spectral features, we modeled the Kast and LRIS spectra at $\delta t<300$\,d, as host-dominated spectra were available for these two instruments\footnote{The +211.5\,d LRIS spectrum was not included in this analysis as it was obtained with a different slit width from the +535.0\,d LRIS observation, which resulted in mismatch in spectral resolution and host continuum shape.}.
We modeled the observed spectra with a combination of blackbody continuum and host galaxy contribution: $f_{\lambda, \rm obs} = A_1 f_{\lambda, \rm BB}+ A_2 f_{\lambda, \rm host}$, where $f_{\lambda, \rm BB}$ is obtained by using the best-fit blackbody parameters (\S\ref{subsec:bbfit}), and $f_{\lambda, \rm host}$ is directly given by the late-time ($\delta t > 480$\,d) spectra. 
$A_1$ and $A_2$ are constants that account for unknown factors (see details in Section 4.2 of \citealt{Yao2022_21ehb}). 
We searched for $A_1$ and $A_2$ that give the minimum $\chi^2$ in rest-frame 4000--7300\,\AA, while excluding the wavelength ranges that contain the Bowen complex (4500--4800\,\AA) and H$\alpha$ (6300--6700\,\AA).

The host and blackbody continuum subtracted spectra are shown in the lower left panel of Figure~\ref{fig:opt_spec}. While the Kast spectra do not show conclusive line features on the blue side (due to the low SNR), in the LRIS spectrum, intermediate-width emission lines around H$\beta$, the Bowen complex (\ion{He}{II} $\lambda$4686 and \ion{N}{III} $\lambda$4640), and possibly H$\gamma$ are clearly present. In addition, as shown in the upper left panel, a broad emission line around H$\alpha$ is evident in all observed spectra at $\delta t<300$\,d. 
Using the classification scheme developed by \citet{vanVelzen2021} and \citet{Hammerstein2023}, the optical spectral subtype of AT2022lri belongs to the TDE-H+He subclass, which is the most common among optically selected TDEs. 

The lower right panel of Figure~\ref{fig:opt_spec} shows the H$\alpha$ profile in the host and continuum subtracted spectra. 
Fitting a Gaussian to each individual spectrum gives the measured full-width half-maximum (FWHM) velocities (from +210.4\,d to +269.4\,d) of $5219 \pm 485\,{\rm km\,s^{-1}}$, $4763\pm 309\,{\rm km\,s^{-1}}$, and $1887 \pm 95\,{\rm km\,s^{-1}}$, indicating that the H$\alpha$ line width becomes narrower at later times. The narrowing of optical lines is common in TDEs \citep{vanVelzen2020, Charalampopoulos2022}.
This might be caused by a drop of gas density in the line-forming region if the line width is mainly determined by electron scattering \citep{Roth2018}. 
The line centers are measured to be at $-1095 \pm 169\,{\rm km\,s^{-1}}$, $-887 \pm 111\,{\rm km\,s^{-1}}$, and $-425 \pm 38\,{\rm km\,s^{-1}}$. 
The slight blueshift by ${\rm few}\times 10^2\,{\rm km\,s^{-1}}$ is also manifested in other spectra of AT2022lri shown in the upper right panel of Figure~\ref{fig:opt_spec}. 
We note that a blueshifted H$\alpha$ emission line has also been previously observed in other TDEs, albeit at much earlier times, such as in the peak-light spectra of ASASSN-14ae \citep{Holoien2014} and AT2018dyb \citep{Leloudas2019}. The blueshifted H$\alpha$ centroid observed in AT2022lri might be related to wide-angle outflows at $\delta t > 200$\,d \citep{Roth2018}.

\section{X-ray Observations and Analysis} \label{sec:xray}

All X-ray spectral fitting was performed with $\chi^2$ statistics.

\subsection{Swift/XRT} \label{subsec:xrt}

\begin{figure*}[htbp!]
	\centering
	\includegraphics[width=0.95\textwidth]{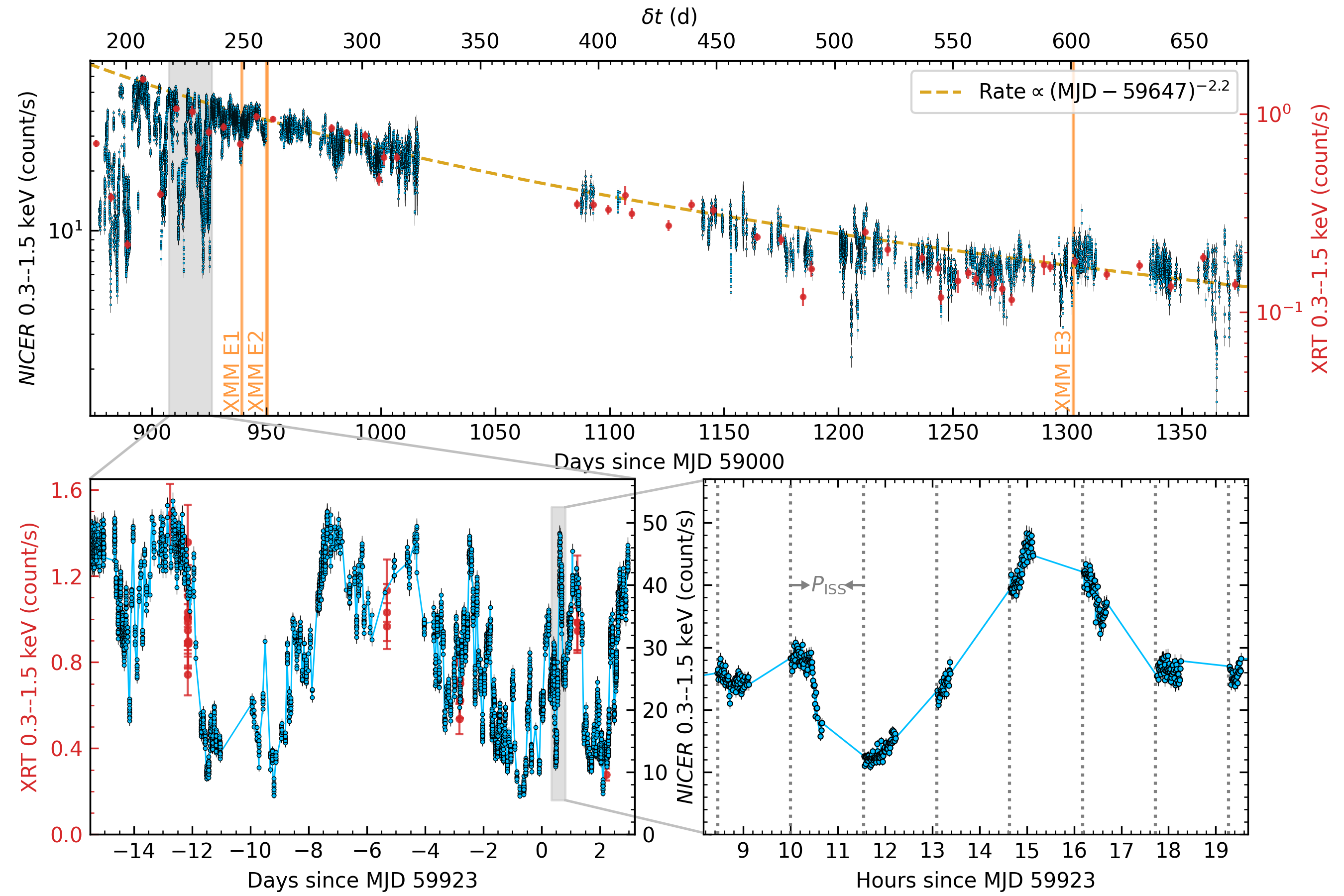}
	\caption{X-ray light curve of AT2022lri.
	\textit{Upper}: The vertical orange bands mark three epochs of \xmm observations. 
	The dashed yellow line shows a power-law decline to guide the eye.
	\textit{Lower}: A zoom-in around MJD 59923 to highlight the variability timescale at $t_{\rm var}\sim1$\,hr. The vertical dotted lines in the lower right panel mark the beginning of each GTI. 
		\label{fig:xray_lc}}
\end{figure*}

All XRT observations were obtained in the photon-counting mode. 
Using an automated online tool\footnote{\url{https://www.swift.ac.uk/user_objects}} \citep{Evans2007, Evans2009}, we generated the XRT 0.3--1.5\,keV light curve (Figure~\ref{fig:xray_lc}), 
and the XRT spectra for all epochs.
All XRT spectra are further grouped with \texttt{ftgrouppha} using the optimal binning scheme \citep{Kaastra2016} and simultaneously ensure at least 25 counts per bin. 

For X-ray spectral modeling, we included the Galactic absorption using the \texttt{tbabs} model \citep{Wilms2000}, with the hydrogen-equivalent column density $N_{\rm H}$ fixed at $1.60\times 10^{20}\,{\rm cm^{-2}}$ \citep{HI4PI2016}.
We also convolved all TDE spectral models with  \texttt{zashift} to account for host redshift.
Since the XRT spectra remain soft, we adopted a standard multi-color disk model (\texttt{diskbb}) to fit the spectra, considering the energy ranges between 0.3\,keV and wherever the net count rate becomes less than 1.5 times the background count rate.

\begin{figure}[htbp!]
\centering
\includegraphics[width=0.95\textwidth]{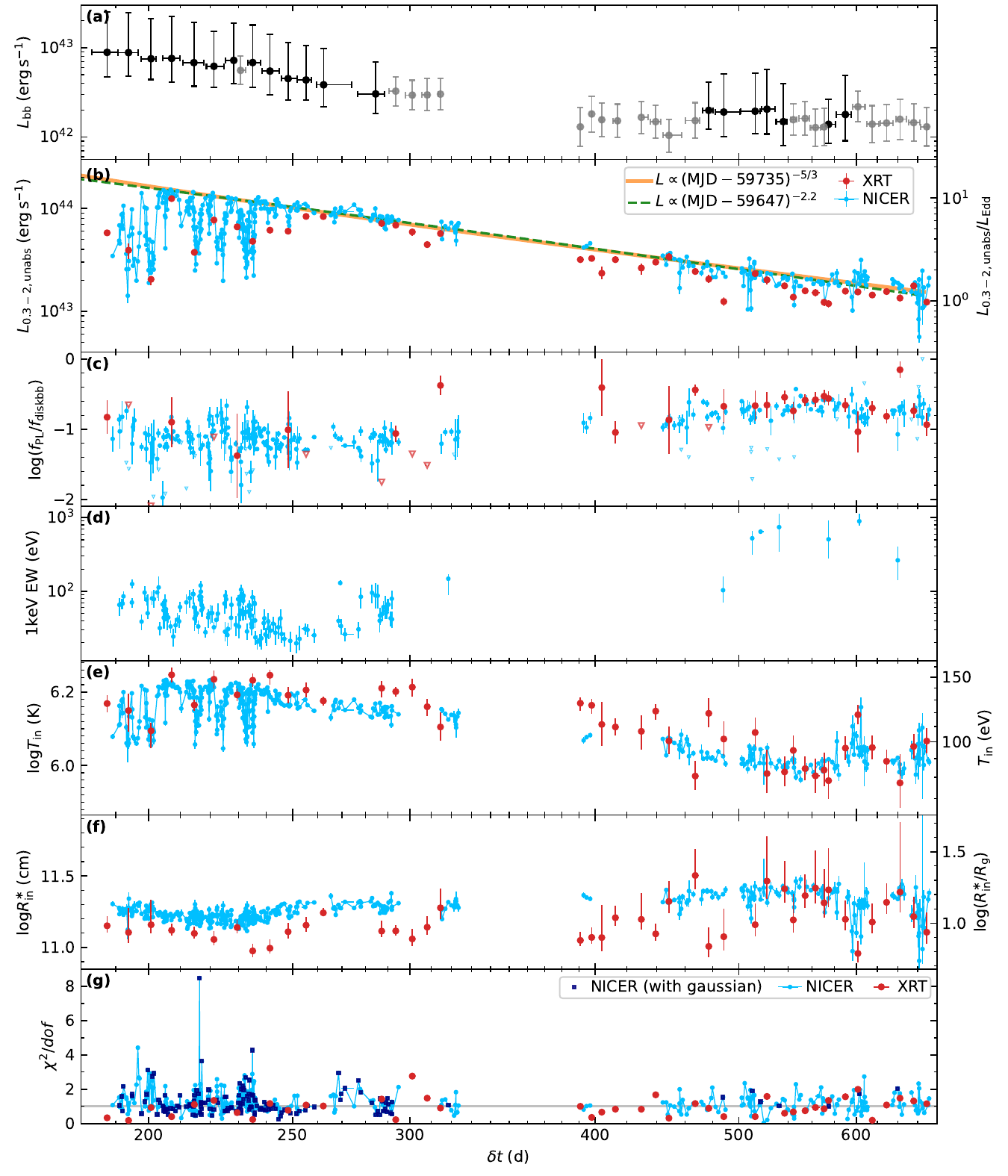}
\caption{Multi-wavelength evolution of AT2022lri.
	Panel (a): blackbody luminosity of the UV/optical emission. Black circles mark epochs where both $T_{\rm bb}$ and $R_{\rm bb}$ are fitted, while gray circles mark epochs where only $R_{\rm bb}$ is fitted (see details in \S\ref{subsec:bbfit}). 
	Panel (b): the unabsorbed rest-frame 0.3--2\,keV X-ray luminosity. 
	Panel (c): the ratio of the 0.3--2\,keV \texttt{powerlaw} to \texttt{diskbb} fluxes.
	Panel (d): equivalent width (EW) of the 1\,keV line, only shown if the model fit with the \texttt{gaussian} component is favored with the BIC criterion (see details in \S\ref{subsubsec:nicer_spec}). 
	Panel (e)--(f): best-fit $T_{\rm in}$ and $R^{\ast}_{\rm in}\equiv R_{\rm in}\sqrt{{\rm cos}i}$ in the \texttt{diskbb} component. 
	Panel (g): fit statistics.
 In the right y axis of panels (b) and (f), we assume $M_{\rm BH}=10^5\,M_\odot$ (see \S\ref{subsec:Mbh}). 
	\label{fig:multiwave_evol}}
\end{figure}

While some of the initial spectra can be well described by the redshifted multi-color disk, at later epochs, the values of $\chi^2/dof$ become large, leaving a hard excess unaccounted for by the model. 
This suggests the existence of non-thermal Comptonized X-ray photons. 
Therefore, we re-modeled the XRT spectra by adding a \texttt{powerlaw} component, which significantly improved the fit. 
If we let the power-law photon index $\Gamma$ to be a free parameter, we find that $\Gamma$ can only be constrained in $\sim\!20$\% of XRT observations, where the best-fit $\Gamma$ ranges from 2.3 to 3.4, with a median at 2.8. 
However, even in these observations, the fractional uncertainties are typically greater than 50\%. Therefore, we re-fitted the spectra with  $\Gamma$ fixed at 2.8. The resulting $\chi^2/dof$ is close to one in most of the cases. 

In Figure~\ref{fig:multiwave_evol}, we show the unabsorbed 0.3--2\,keV X-ray luminosity ($L_{\rm X}$, panel b), the ratio of 0.3--2\,keV fluxes in the \texttt{powerlaw} to \texttt{diskbb} components ($f_{\rm PL}/f_{\rm diskbb}$, panel c), the inner disk temperature ($T_{\rm in}$, panel e), the apparent inner disk radius times square root of ${\rm cos} i$, where $i$ is the system inclination ($R^{\ast}_{\rm in}\equiv R_{\rm in} \sqrt{\cos i}$, panel f), and the $\chi^2/dof$ values (panel g). 
As can be seen, from $\delta t \sim\!190$\,d to $\delta t\sim\!570$\,d, $f_{\rm pl}/f_{\rm diskbb}$ has gradually increased from  
$\sim\!6$\% to 
$\sim\!25$\%. 
While from $\delta t\sim\!570$\,d to $\delta t\sim\!680$\,d it seems to have slightly decreased again to $\sim\!16$\%. 

\subsection{XMM-Newton}

We obtained three epochs of observations with the \xmm telescope (indicated by the vertical orange lines in Figure~\ref{fig:xray_lc}) through our GO and Directors Discretionary Time (DDT) programs. An observing log is given in Appendix~\ref{sec:table_log} (Table~\ref{tab:xmm-log}). Below we describe analysis of data from two instruments: the European Photon Imaging Camera (EPIC) which has the pn \citep{Struder2001} and MOS chips, and the Reflection Grating Spectrometer (RGS; \citealt{denHerder2001}).

\subsubsection{EPIC Analysis} \label{subsubsec:epic}
Standard reduction procedures for the EPIC camera were employed, as detailed in \citet{Guolo2024}. Prior to extracting X-ray spectra, we had to account for effects of photon pile-up, which is prevalent in soft sources with high X-ray flux. 

For the XMM\,E1 observation performed in the ``Full Frame'' mode, the pile-up was so extreme that even excluding a large portion of the PSF did not correct all the pile-up spectral distortions. We therefore decided to not use the EPIC camera data for this observation. 

The XMM\,E2 and XMM\,E3 observations were performed in the ``Small Window'' mode, and pile-up correction is feasible. In both observations we considered an annular source region with a fixed outer radius of $32^{\prime\prime}$ 
centered on the source coordinates. For XMM\,E2 (XMM\,E3) we chose an inner exclusion radii of $11^{\prime\prime}$ ($3^{\prime\prime}$) 
after quantifying the pile-up in each case by following the procedure outlined on \xmm's data analysis page\footnote{\url{https://www.cosmos.esa.int/web/xmm-newton/sas-thread-epatplot}}.
A circle with $r_{\rm bkg}=45^{\prime\prime}$ 
outside the source area, but within the same CCD, was chosen to generate the background spectrum. Using the \texttt{evselect} task, we only retained patterns that correspond to single events. 
Since the pile-up was even more extreme in MOS1 and MOS2, and that the pn instrument has better sensitivity, we only analyzed the pn data.
The resulting XMM\,E2 and XMM\,E3 pn spectra had, respectively, $\sim\!2\times10^5$ and $\sim\!8\times10^4$ background subtracted counts, both were binned using the optimal binning criteria \citep{Kaastra2016} while ensuring that each bin has at least 25 counts. 

\begin{figure}[htbp!]
	\centering
	\includegraphics[width=0.8\columnwidth]{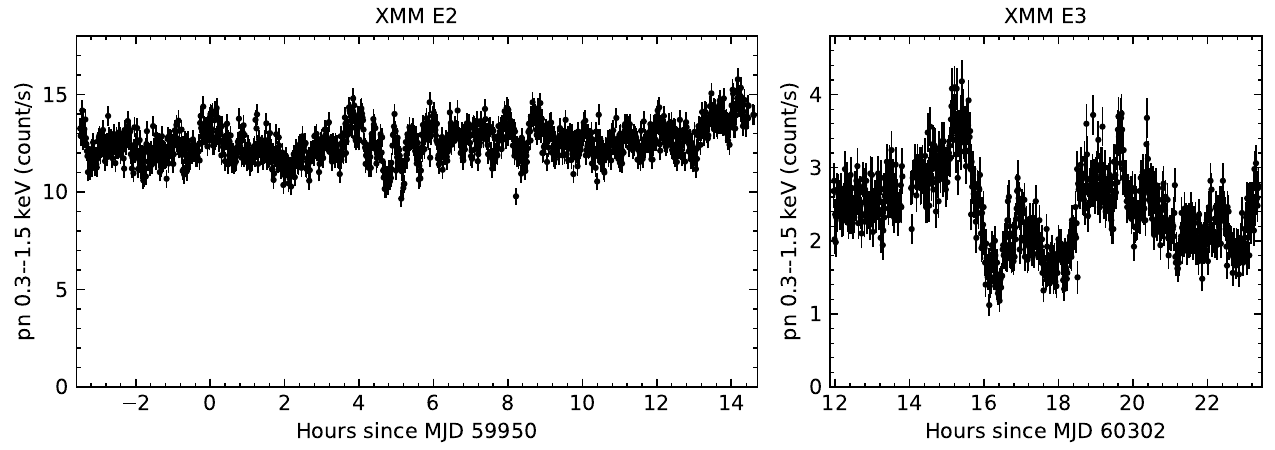}
	\caption{\xmm/EPIC light curve of AT2022lri in the second (left) and third (right) epochs. 
 Bins with background flaring have been masked out. 
		\label{fig:xmm_lc}}
\end{figure}

Figure~\ref{fig:xmm_lc} shows the 50\,s-binned light curve of the XMM\,E2 and XMM\,E3 observations. In XMM\,E2, no strong variability was observed. While in XMM\,E3, we observe a flux variation by a factor of $\approx\!3.7$ within $\approx\!44$\,min. 

\begin{figure}[htbp!]
	\centering
	\includegraphics[width=0.45\textwidth]{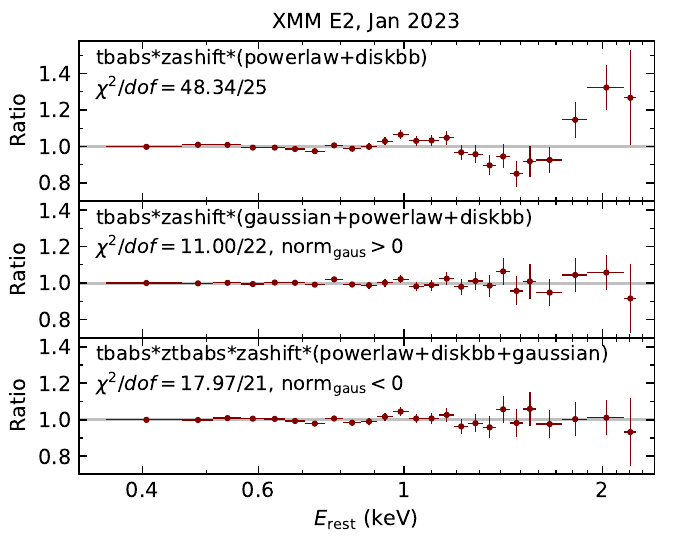}
	\includegraphics[width=0.45\textwidth]{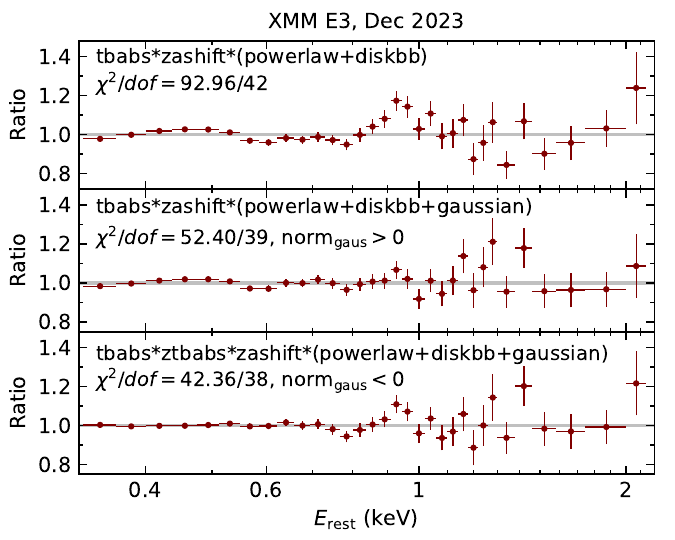}
	\caption{Ratio of the XMM\,E2 (left) and XMM\,E3 (right) EPIC/pn data fit to various models.
		\label{fig:xmm_pn_spec}}
\end{figure}

For EPIC spectral fitting, we consider data between 0.3\,keV and wherever the net count rate becomes less than 1.5 times the background count rate, which gives 2.2\,keV for XMM\,E2 and 2.1\,keV for XMM\,E3.
Since a combination of \texttt{diskbb} and \texttt{powerlaw} provides acceptable fits to the XRT data (\S\ref{subsec:xrt}), we started from this continuum model, allowing the disk temperature, \texttt{diskbb} normalization, $\Gamma$, and \texttt{powerlaw} normalization to be free parameters.
In both observations, we obtained poor fits, with $\chi^2/dof=48.34/25$ in XMM\,E2 and $\chi^2/dof=92.96/42$ in XMM\,E3 (see Figure~\ref{fig:xmm_pn_spec}, top rows). 
Changing the \texttt{diskbb} component to other thermal models such as \texttt{tdediscspec} \citep{Mummery2021} and \texttt{bbody} or changing the  \texttt{powerlaw} component to other Comptonization models such as \texttt{simpl} \citep{Steiner2009} or \texttt{cutoffpl} do not fit the residual.

To improve the fit, we first tried adding a \texttt{gaussian} emission component, allowing the line center ($E_{\rm line}$) to vary between 0.8 and 1.2\,keV and the normalization (norm$_{\rm gaus}$) to be positive. 
This significantly improved the fits (see Figure~\ref{fig:xmm_pn_spec}, middle rows).
The best-fit $E_{\rm line}$ is capped at the minimum value of 0.8\,keV in XMM\,E2, and $E_{\rm line}=0.90_{-0.04}^{+0.03}$\,keV in XMM\,E3.

Alternatively, we also tried adding a \texttt{gaussian} absorption component, allowing $E_{\rm line}$ to vary between 1.0 and 1.8\,keV and norm$_{\rm gaus}$ to be negative. In this scenario, we found that the models systematically overpredict the flux at $\lesssim 0.4$\,keV. Therefore, we added an additional component of neutral absorption at the host redshift (\texttt{ztbabs}). 
This also gives acceptable fits (see Figure~\ref{fig:xmm_pn_spec}, bottom rows).
In XMM\,E2, the best-fit $E_{\rm line}=1.35_{-0.09}^{+0.05}$\,keV and $N_{\rm H, host}=1.46_{-0.64}^{+0.64}\times10^{20}\,{\rm cm^{-2}}$.
In XMM\,E3, the best-fit $E_{\rm line}=1.16_{-0.09}^{+0.09}$\,keV and $N_{\rm H, host}=2.80_{-0.49}^{+0.52}\times10^{20}\,{\rm cm^{-2}}$.

Since the spectral residuals after adding an emission or absorption line are not particularly strong, we did not apply other physically motivated models to these observations. 

\subsubsection{RGS Analysis} \label{subsubsec:rgs}

\begin{figure}[htbp!]
	\centering
	\includegraphics[width=0.7\columnwidth]{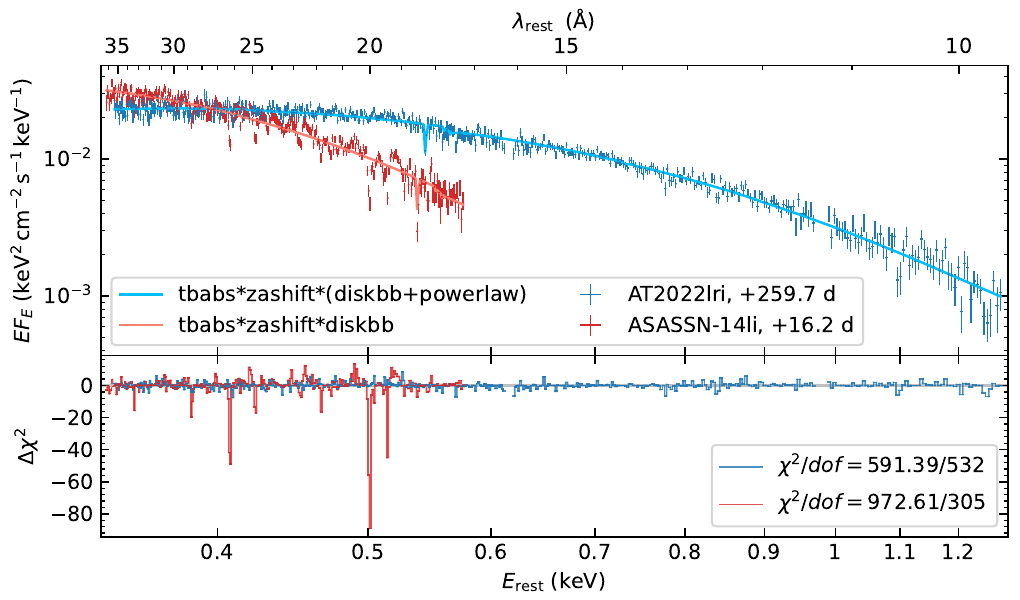}
	\caption{A comparison between the RGS spectrum of AT2022lri to that of ASASSN-14li \citep{Miller2015}. While strong narrow absorption lines were seen in ASASSN-14li, the spectrum of AT2022lri can be well described by the continuum model. 
		\label{fig:rgs_spec}}
\end{figure}

In the XMM\,E2 visit, the long exposure time and high X-ray flux of AT2022lri allowed for a high SNR RGS spectrum. 
Data reduction follows procedures adopted by \cite{Miller2015}, which presented the RGS spectrum of the TDE ASASSN-14li.
The \texttt{rgsproc} routine was used to generate RGS spectral files from the source, background spectral files, and instrument response files. 
The spectrum was binned by a factor of five for clarity.

We model the wavelength range where the net count rate is greater than 1.5 times the background count rate, which gives 9.9--37.3\,\AA. A combination of \texttt{diskbb} and \texttt{powerlaw} gives an acceptable fit, with $\chi^2/dof=591.39/532$. 
No strong absorption lines were evident in the RGS spectrum (see Figure~\ref{fig:rgs_spec}). 
This suggests two possibilities: either there is no low-velocity outflow obscuring the inner disk, or the outflow's column density and ionization state are at levels where the lines are too faint to be observed effectively. 
The potential presence of very weak lines is beyond the scope of this manuscript. 

\subsection{NICER} \label{subsec:nicer}
\nicer is an X-ray telescope on the International Space Station (ISS), which has an orbital period of $P_{\rm ISS}=92.9$\,min. The X-ray timing instrument (XTI) contains 56 X-ray ``concentrators'' \citep{Okajima2016} and the associated focal plane module (FPM) detectors. 
High-cadence \nicer X-ray observations were obtained as part of our GO programs (PI: Y. Yao, IDs 5035, 6078; PI: T. Wevers, ID 6124). 
We note that although the \nicer GO program asked for a lower cadence (3\,ks per two days), the \nicer team increased cadence based on initial quick look data.

Within the \nicer field of view (FoV) of $30\,{\rm arcmin}^2$, there exists no bright X-ray objects close to AT2022lri (see details in Appendix~\ref{subsec:Xray_image}). 
Data reduction was performed using \texttt{HEASoft} version 6.32.1, which contains the \nicer Data Analysis Software (\texttt{nicerdas}) version 11a. We processed the data with \texttt{nicerl2} following the standard pipeline.

\subsubsection{Light Curves} \label{subsubsec:nicer_lc}
Using \texttt{nicerl3-lc}, we extracted light curves in the 0.3--1.5\,keV range
with a time bin of 32\,s. 
The extracted light curves are normalized to an effective area with 52 detectors. 
Additional screening criteria were applied to improve the light curve quality. 
First, we removed time bins with high background noise by requiring that the 13--15\,keV (background dominated) count rate is less than 0.5\,\cps. 
Next, we set the minimum acceptable fractional exposure to be 0.7 by removing time bins with \texttt{FRACEXP<0.7}. 
Finally, we found that some obsIDs were affected by local obstructions due to ISS hardware near to the \nicer FoV\footnote{See details of this issue at \url{https://heasarc.gsfc.nasa.gov/docs/nicer/analysis_threads/iss_obstruction/}.}. In our observations, such obscurations typically cause the egress (ingress) of a ``dipping'' behavior at the beginning (end) of certain GTIs. We visually inspected the light curves, and excluded dips at GTI edges that were associated with lower values of \texttt{ST\_STARS} and \texttt{ST\_OBJECTS} in the filter file.

The top panel of Figure~\ref{fig:xray_lc} shows the \nicer 0.3--1.5\,keV light curve of AT2022lri. The light curve at $\delta t \lesssim 260$\,d exhibits large-amplitude variability on top of a decaying trend, as indicated by the dashed yellow line. 
The lower panels of Figure~\ref{fig:xray_lc} zoom in around MJD 59923 to highlight the observed short-timescale X-ray variability. For the highest cadence \nicer observations (e.g., shown in the lower right panel), the separation between consecutive GTIs is $P_{\rm ISS}$. Intra-GTI variability is clearly observed. 
The flux can change by a factor of $\approx\!2$ in $\approx \! 30$\,min, by a factor of $\approx\!5$ in $\approx\!0.5$\,d, and by a factor of $\approx\!8$ in $\approx 1$\,d.

We ran a periodicity search on the \nicer light curve by computing the Lomb–Scargle \citep{VanderPlas2018} and the multi-harmonic analysis of variance (MHAOV; \citealt{Schwarzenberg-Czerny1996}) 
periodograms, using a frequency grid from 0.002\,d$^{-1}$ to 3\,d$^{-1}$. 
For the MHAOV method, we employed a Fourier series model of five harmonics. 
The periodograms were computed with both the full \nicer light curve and sub-chucks of the light curve across the evolution. 
No significant periodicity aside from $P_{\rm ISS}$ or its harmonics was found.

\subsubsection{Basic Spectral Modeling} \label{subsubsec:nicer_spec}
While most previous TDE analysis utilizing \nicer data focused on the obsID-grouped spectra, we note that AT2022lri exhibits strong variability within some obsIDs. For example, the lower right panel of Figure~\ref{fig:xray_lc} shows a subset of data within obsID 5535022401. In such cases, spectral properties derived from the obsID-grouped data might not provide appropriate characterization of the X-ray properties. This issue is more severe during phases where higher amplitude X-ray variability was observed. 

To mitigate this effect, we selected time boundaries for each spectrum to ensure that (i) the fractional count rate variation within each spectrum is less than 30\%, and (ii) GTIs separated beyond a gap of 2\,d are divided into different spectra.
The time files were created with \texttt{maketime}. 
Event filtering and spectral extraction were then performed with \texttt{niextract-events} and \texttt{nicerl3-spect}, respectively.
Background spectra were created with the \texttt{3c50} model \citep{Remillard2022}. 
By default, \texttt{nicerl3-spec} adds a systematic uncertainty of $\sim\!1.5$\%.
Using \texttt{ftgrouppha}, we grouped the spectra using the \citet{Kaastra2016} optimal binning method and simultaneously ensure a minimum of 25 counts per bin. 

A total of 479 time-resolved spectra were generated. 
We first determined the energy range for spectral fitting by requiring that the net count rate is above 1.5 times the background count rate.
We capped the lower boundary ($E_{\rm lower}$) to be $\geq 0.25$\,keV and the upper boundary ($E_{\rm upper}$) to be $\leq10.0$\,keV. 
Next, we removed spectra with high background levels, selected if the 3--8\,keV net count rate is greater than 0.8\,\cps or if the 0.25--0.35\,keV background count rate is greater than 0.5\,\cps.
We also discarded spectra where the number of spectral bins between $E_{\rm lower}$ and $E_{\rm upper}$ is less than 8. 
A total of 429 spectra survived these cuts. 

\begin{figure}[htbp!]
	\centering
	\includegraphics[width=0.9\columnwidth]{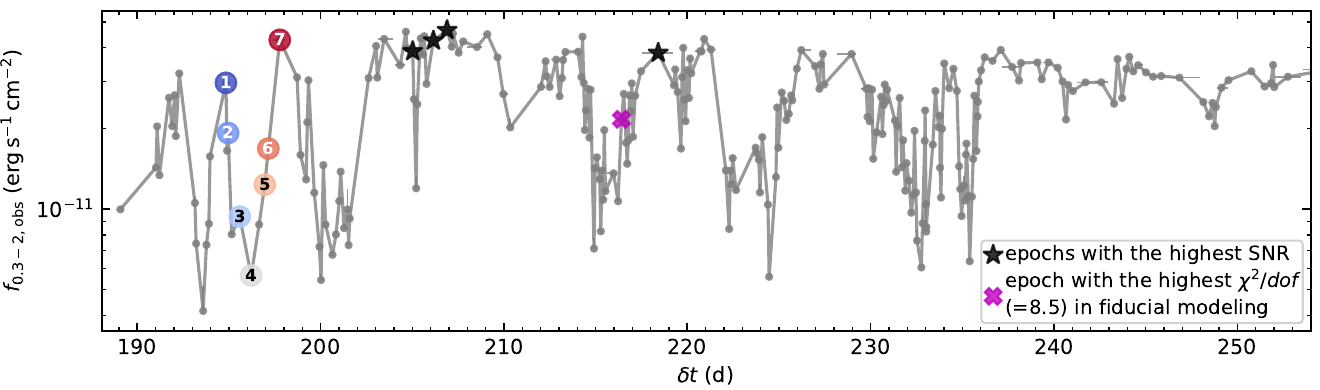}
	\caption{\nicer observed 0.3--2\,keV light curve of AT2022lri at $\delta t < 254$\,d. The black asterisks mark four observations with high-quality data ($E_{\rm upper}>2.1$\,keV, net counts $>5\times 10^4$) selected for an initial basic spectral modeling. The magenta cross mark one observation selected for detailed modeling (see Figure~\ref{fig:nicer_highstat_specs}). The large circles mark the assigned indices of seven observations during one X-ray dip (see Figure~\ref{fig:nicer_drop1_sp}).
		 \label{fig:nicer_early_lc}}
\end{figure}

To obtain a general sense of the \nicer spectral shape, we first looked at four spectra with the highest SNR by selecting those with broad energy ranges ($E_{\rm upper}>2.1$\,keV) and total net counts $>5\times 10^4$ (see the black asterisks in Figure~\ref{fig:nicer_early_lc}). We found that in all four observations, when fitted with a combination of \texttt{diskbb} and \texttt{powerlaw}, there exist residual features, which can be modeled with either an emission line at $\sim\!1$\,keV or an absorption line at $\sim\!1.3$\,keV, similar to what was found in the EPIC spectra (see \S\ref{subsubsec:epic}).

For the purpose of a simple phenomenological modeling, we then decided to fit all time-resolved \nicer spectra with two models (1) \texttt{tbabs*zashift*(powerlaw+diskbb)}, and (2) \texttt{tbabs*zashift*(gaussian+powerlaw+diskbb)}, where the line width and normalization are allowed to be free (norm$_{\rm gaus}>0$), and $E_{\rm line}$ is fixed at 1\,keV.
The fact that $E_{\rm upper}\sim\!1$\,keV in most \nicer spectra precludes constraints on the shape of the hard component. 
Therefore, in both models, we fixed $\Gamma$ at 2.8.
We assess the goodness of fit by computing the model Bayesian information criterion (BIC): ${\rm BIC} = k \cdot {\rm ln}(N) -2 {\rm ln}\mathcal{ L} = k \cdot{\rm ln}(N) + \chi^2 + {\rm constant}$, where $k$ is the number of free parameters, $N$ is the number of spectral bins, and $\mathcal{L}$ is the maximum of the likelihood function. Following the \citet{Raftery1995} guidelines, we select the fitting results from model (2) if its BIC value is more than 6 smaller than that in model (1).

\begin{figure}[htbp!]
\centering
\includegraphics[width=0.95\columnwidth]{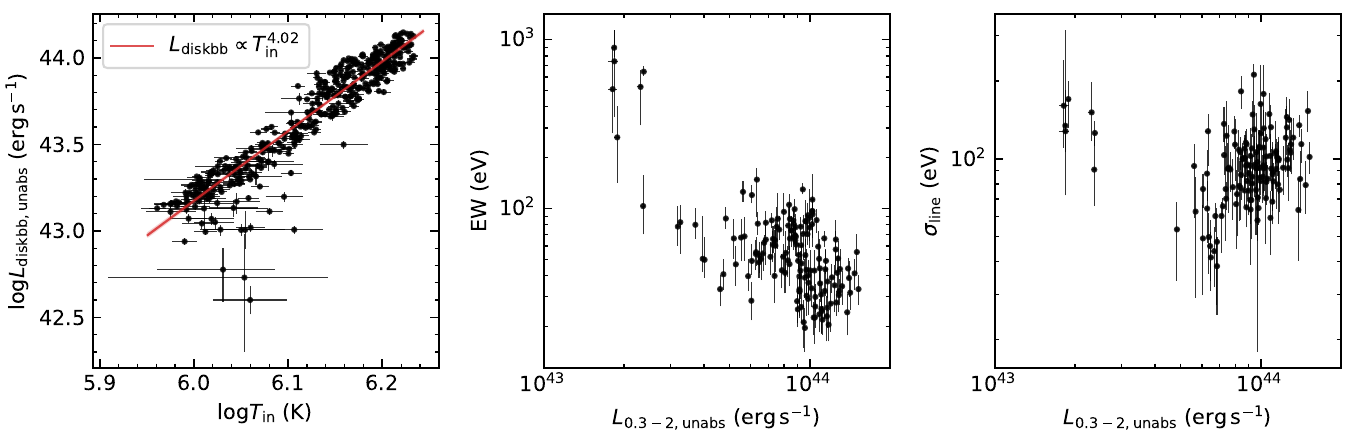}
\caption{Model parameters in basic \nicer spectral fitting. 
\textit{Left}: inner disk temperature vs. the unabsorbed rest-frame 0.3--2 \,keV luminosity in the \texttt{diskbb} component.  
\textit{Middle}: equivalent width of the 1\,keV line vs. the unabsorbed rest-frame 0.3--2\,keV luminosity $L_{\rm 0.3-2, unabs}$, only plotted for observations where the model fit with the \texttt{gaussian} component is favored with the BIC criterion. 
\textit{Right}: line width $\sigma_{\rm line}$ of the \texttt{gaussian} component vs. $L_{\rm 0.3-2, unabs}$.
\label{fig:nicer_xpars_corr}}
\end{figure}

The fitting results are presented in Figure~\ref{fig:multiwave_evol} and Figure~\ref{fig:nicer_xpars_corr}. 
The unabsorbed rest-frame 0.3--2\,keV X-ray luminosity ($L_{\rm 0.3-2, unabs}$) evolution shows short-timescale dips on top of a general long-term declining trend, which we call the ``envelope'' of the X-ray light curve.
Assuming that the ``envelope'' follows a power-law $\propto ({ t-  t_{\rm disr}})^{-\alpha}$, we found that the best-fit  $t_{\rm disr}=59735\pm 3$ (in MJD) if $\alpha = 5/3$ 
and $t_{\rm disr} = 59647\pm4$ if $\alpha = 2.2$. 
We adopt the fit with $\alpha = 2.2$ as the disruption epoch $t_{\rm disr}$ should precede the first optical detection epoch (see Figure~\ref{fig:lc_uvopt} and \S\ref{subsec:opt_phot}).
The observed peak luminosity in the ``envelope'' reaches a maximum of $1.5\times 10^{44}\,{\rm erg\,s^{-1}}$ at $\delta t \approx\!205$\,d and falls to $1.5\times 10^{43}\,{\rm erg\,s^{-1}}$ at $\delta t \approx\!670$\,d.

Throughout the X-ray evolution, the inferred $R^{\ast}_{\rm in}$ parameter remains roughly constant at $\sim\!2\times 10^{11}$\,cm. 
Both the long-term X-ray flux decline trend and the X-ray dips at $\delta t \lesssim 240$\,d seem to be correlated with changes in the inner disk temperature $T_{\rm in}$. Fitting a function of the type log$(L_{\rm diskbb}/{\rm erg\,s^{-1}}) = a + b \times {\rm log}(T_{\rm in}/{\rm K}) + \epsilon$ yields a slope of $b = 4.02\pm0.05$ and an intrinsic scatter of $\epsilon=0.08$, as shown in the left panel of Figure~\ref{fig:nicer_xpars_corr}. 

In observations where model (2) is favored with the BIC criterion, we found that the equivalent width (EW) of the 1\,keV emission lines varies between $\approx \!20$\,eV to $\approx \! 800$\,eV (see panel d of Figure~\ref{fig:multiwave_evol} and the middle panel of Figure~\ref{fig:nicer_xpars_corr}). The line width follows $30\,{\rm eV}\lesssim \sigma_{\rm line}\lesssim 200$\,eV (see the right panel of Figure~\ref{fig:nicer_xpars_corr}). 

As shown in the bottom panel of Figure~\ref{fig:multiwave_evol}, in some epochs, the selected model does not provide a fit with $\chi^2/dof<2$, indicating that additional spectral features (other than a flux excess with a Gaussian profile at 1\,keV) are present in the data. 


\subsubsection{Modeling Additional Spectral Features} \label{subsubsec:nicer_1keV_line}

To investigate the shape of additional spectral features, we first focused on one observation with the highest value of $\chi^2/dof$ ($=8.5$) in the phenomenological spectral modeling (\S\ref{subsubsec:nicer_spec}). With total net counts $=7.5\times 10^4$, this observation (${\rm MJD}=59905.498$, marked with a magenta cross in Figure~\ref{fig:nicer_early_lc}) has a high SNR. 

\begin{figure}[htbp!]
\centering
\includegraphics[width=0.8\columnwidth]{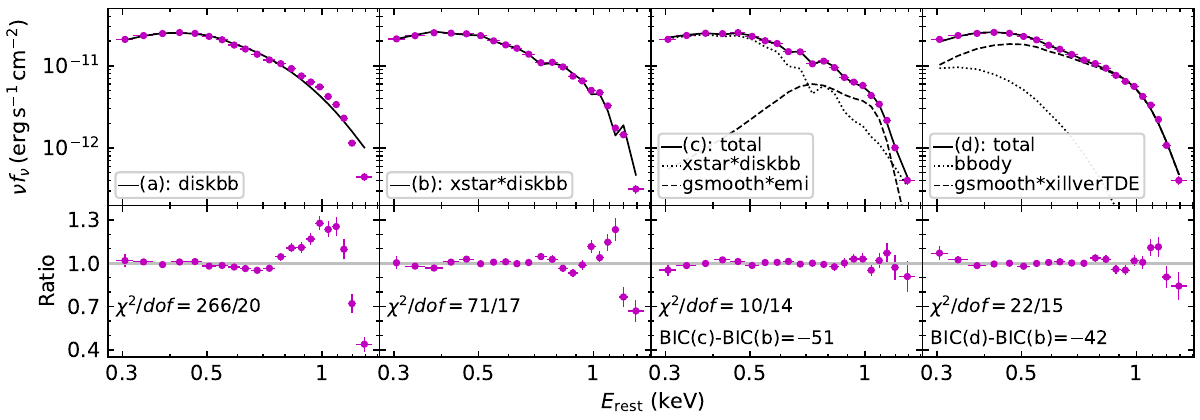}
\caption{Spectral modeling results of the \nicer observation on MJD 59905.498 --- the one with the highest $\chi^2/dof$ in the initial analysis.  
From left to right, we show the data, best-fit models, and residuals under four different models. Models (c) and (d) give acceptable fits.
\label{fig:nicer_highstat_specs} }
\end{figure}

As a start, we fitted the data with our fiducial model of \texttt{tbabs*zashift*diskbb}. A hard \texttt{powerlaw} component was found not to be needed for this observation, and was therefore not included.
As can be seen in the the bottom left panel of Figure~\ref{fig:nicer_highstat_specs}, the fiducial model gives very strong residuals. Adding a \texttt{gaussian} emission or absorption component significantly improves the fit, but neither gives $\chi^2/dof<3$. 

We attempted to improve the fit by adding absorption component(s) on top of our fiducial continuum model.
We began by implementing neutral absorption (i.e., \texttt{tbabs}) at the host redshift. 
The best-fit model did not improve the fit, with the best-fit $N_{\rm H}$ being close to zero.
We also tried partial covering of neutral absorption (\texttt{pcfabs}), where a fraction of the X-ray source is seen through a neutral absorber, while the rest is assumed to be observed directly. This approach did not lead to an improved fit either.

Next, we tested the idea of absorption from an ionized absorber. 
We performed photoionization absorption modeling using an absorption grid generated by \texttt{XSTAR} \citep{Kallman2001}, assuming an ionizing spectrum that follows the best-fit \texttt{diskbb+powerlaw} continuum at $\delta t = 218.4$\,d when there was no X-ray dipping (i.e., $T_{\rm in}=150$\,keV, $\Gamma=3.5$, shown as a black asterisk in Figure~\ref{fig:nicer_early_lc}). 
The absorption model has three free parameters: the hydrogen column density of the ionized absorber $N_{\rm H}$, the ionization state log$\xi$, and the redshift of the absorber $z$\footnote{Throughout the rest of this paper, we only report the outflow velocity inferred using the relativistic Doppler formula, $1+z=\sqrt{(1+v/c)/(1-v/c)}$.}. 
Given the relative breadth of the spectral features and the limited energy resolution, we generated a model grid with a turbulent velocity of $v_{\rm turb}=10^4\,{\rm km\,s^{-1}}$.
Applying \texttt{tbabs*zashift*xstar*diskbb} does not fully account for the residuals, shown as model (b) in Figure~\ref{fig:nicer_highstat_specs}. 

After that, we investigated if the fitting results can be further improved by adding emission from the ionized absorbers by using a model of \texttt{tbabs*zashift*(xstar*diskbb+gsmooth*emi)}. 
Similar models have been applied to AGN \citep{Nardini2015, Tombesi2015, Laurenti2021}
and TDE \citep{Kara2018}. 
Here \texttt{emi} is an \texttt{XSTAR} emission table calculated with the same parameters as for the absorption table. It has four parameters: the column density $N_{\rm H}$, the ionization state log$\xi$, the redshift of the emission component $z$, and the normalization norm$_{\rm emi}$. 
We tied the first two parameters to be the same as that in the \texttt{xstar} absorption component, and let the last two parameters to be free.
Note that $z$ in \texttt{emi} can be different from $z$ in \texttt{xstar}, because the geometry of the outflow influences it. 
For instance, for a spherically symmetric outflow, we would expect a P-Cygni profile with blueshifted absorption along the line of sight and broad redshifted emission due to the other parts of the outflow not along the line of sight. If there is a disk occulting the back outflow, we would expect both emission and absorption to be blueshifted, but with different shifts due to the fact that emission comes from different directions. Roughly, the width of the emission should be comparable to the absorption outflow velocity. 
The \texttt{gsmooth} component is added because for a wide-angle outflow we expect emission to come from different regions of the outflow, with different projections along our line of sight. Therefore, we could get a broader emission profile after integrating the emission through the wind geometry. 
The fitting result is shown as model (c) in Figure~\ref{fig:nicer_highstat_specs}, which gives $\chi^2/dof = 10.27/14$ and a BIC value much smaller than that of model (b). 
The best-fit parameters are: $N_{\rm H} = 2.37_{-0.61}^{+0.86} \times 10^{21}\,{\rm cm^{-2}}$, log($\xi$/erg\,cm\,s$^{-1})=2.98_{-0.03}^{+0.17}$, $v_{\rm out}/c=-0.110_{-0.013}^{+0.014}$ (in \texttt{xstar}), $T_{\rm in}=116.9_{-2.7}^{+2.4}$\,eV, norm$_{\rm diskbb}=20.75_{-2.14}^{+2.86}\times10^3$, $\sigma=0.089_{-0.007}^{+0.008}$\,keV, $v_{\rm out}/c=-0.134_{-0.009}^{+0.010}$ (in \texttt{emi}), and norm$_{\rm emi} = 8.98_{-1.42}^{+2.35} \times 10^{-3}$.
We conclude that the model with both emission and absorption from the wind are favored over the one with only absorption.


Additionally, we test a newly developed physically motivated reflection model called \texttt{xillverTDE}.
While classic reflection models (e.g., \texttt{xillver}, \texttt{xillverCP}; \citealt{Garcia2010, Garcia2013}) are suitable for AGN and low-mass black hole binaries in the hard-state, the power-law irradiation continuum in these models does not apply to thermally dominated X-ray spectra. 
In \texttt{xillverTDE}, the continuum is defined by a single-temperature blackbody spectrum with $0.03\,{\rm keV}<kT_{\rm bb}<0.3$\,keV, appropriate for the inner regions of a geometrically thick disk in TDEs \citep{Masterson2022}. 
\texttt{xillverTDE} has seven free parameters: 
the blackbody temperature $kT_{\rm bb}$, 
the iron abundance with respect to its solar value $A_{\rm Fe}$, 
the ionization parameter at the surface of the disk $\xi$, 
the gas density $n$, 
the inclination $i$, 
the redshift $z$ of the materials that reprocess the irradiation continuum, 
and the normalization norm$_{\rm refl}$.

We applied the reflection model of \texttt{tbabs*zashift*(gsmooth*xillverTDE+bbody)}, fixing $A_{\rm Fe}=1$ and $i=45^{\circ}$ (as in \citealt{Masterson2022}) since the fit is not sensitive to these parameters. 
We tied the blackbody temperature in the \texttt{xillverTDE} and \texttt{bbody} components to be the same. The \texttt{xillverTDE} component is further convolved with \texttt{gsmooth} to account for velocity broadening, which was found to exhibit a more symmetric profile in super-Eddington accretion flows \citep{Thomsen2022_Xrayline}.
The fitting result, with $\chi^2/dof=22.40/15$ and named as model (d), is shown in the right panels of Figure~\ref{fig:nicer_highstat_specs}. The best-fit parameters are: 
$kT_{\rm bb}=67.3_{-5.0}^{+3.8}$\,eV, 
norm$_{\rm bb}=27.4_{-1.8}^{+1.4}\times10^{-5}$, 
$\sigma=0.145_{-0.014}^{+0.004}$\,keV (in \texttt{gsmooth}),
log($\xi$/erg\,cm\,s$^{-1})=3.47_{-0.10}^{+0.01}$,
log$(n$/cm$^{-3})=17.99_{-0.44}^{+0.01}$,
$v_{\rm out}/c=-0.262_{-0.028}^{+0.021}$,
and norm$_{\rm refl}=19.0_{-1.3}^{+12.1} \times 10^{-13}$.

We conclude that both model (c) and model (d) give acceptable fits. According to BIC, for this observation, the reflection model is not favored over modeling with an ionized absorber.

\subsubsection{Modeling of an X-ray Dip} \label{subsubsec:dip}

To investigate the evolution of spectral parameters during the rapid X-ray variability, we then focused on data at $\delta t\sim\!196$\,d and selected seven observations (hereafter referred as sp1--sp7) that sample one X-ray dip for a detailed analysis (see Figure~\ref{fig:nicer_early_lc} for epochs of these observations). Another goal of this step is to verify if the absorption/emission and reflection models that provide acceptable fits in \S\ref{subsubsec:nicer_1keV_line} can be applied in other observations. 
A summary of the fit statistics of our models is given in Table~\ref{tab:model_compare_dip}. 
Below we describe the fitting procedures.

\begin{deluxetable}{clcc}[htbp!]
\tablecaption{Statistics of various models fitted to the seven observations at $\delta t\sim 196$\,d.\label{tab:model_compare_dip}}
\tabletypesize{\footnotesize}
\tablehead{
    \colhead{Index} &
    \colhead{Model}             &
     \colhead{$\chi^2/dof$}   &
     \colhead{$\Delta$BIC}
}
\startdata
(A) & Fiducial model: \texttt{tbabs*zashift*continuum}& 265.13/115 & 0 \\
(B) & \texttt{tbabs*zashift*xstar*continuum} & 151.01/106 & $-70.2$\\
(C) & \texttt{tbabs*zashift*(xstar*continuum+gsmooth*emi)} & 102.83/97 & $-74.5$\\
(D) & \texttt{tbabs*zashift*(gsmooth*xillverTDE+continuum)} & 111.20/104 & $-80.8$\\
     \enddata
\tablecomments{$\Delta$BIC is the difference of BIC values between each model and model (A). In models (A), (B), and (C), \texttt{continuum=diskbb} for sp1--sp6, and \texttt{continuum=(diskbb+powerlaw)} for sp7. 
In model (D), \texttt{continuum=bbody} for sp1--sp6, and \texttt{continuum=(bbody+powerlaw)} for sp7.}
\end{deluxetable}

\begin{figure}[htbp!]
\centering
\includegraphics[width=\columnwidth]{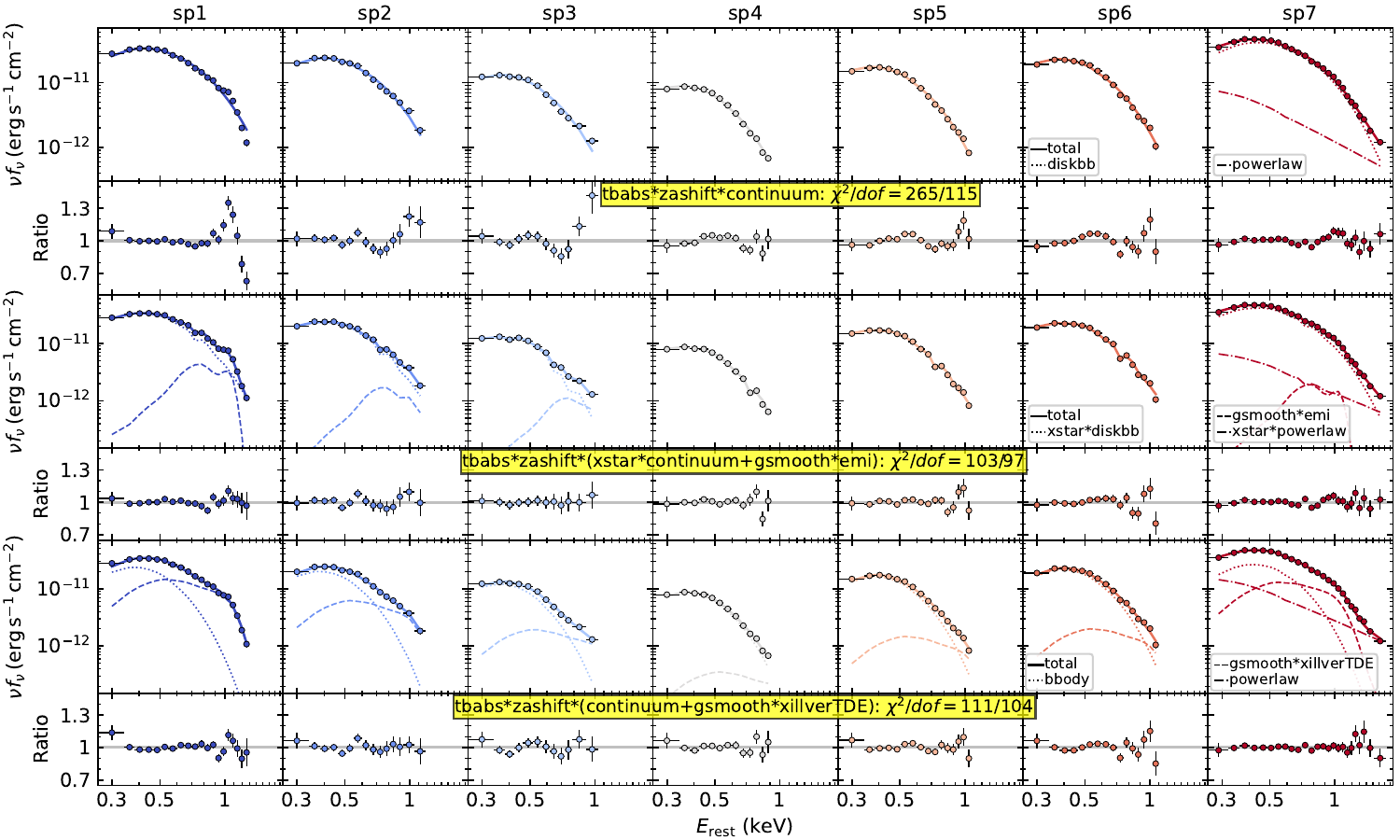}
\caption{Seven observations around $\delta t \sim \! 196$\,d jointly modeled with three different models. We show the data, best-fits, and residuals for the fiducial continuum model (A) in the top two rows, the absorption+emission model (C) in the middle two rows (see Table~\ref{tab:nicer_drop1_modC1} for the best-fit parameters), and the reflection model (D) in the bottom two rows (see Table~\ref{tab:nicer_drop1_modD} for the best-fit parameters). 
\label{fig:nicer_drop1_sp}}
\end{figure}

Before applying complex models, we first started by individually fitting the seven spectra with \texttt{tbabs*zashift*continuum}, where \texttt{continuum=diskbb} for sp1--sp6, and \texttt{continuum=(diskbb+powerlaw)} for sp7\footnote{Since $E_{\rm upper}\leq 1.3$\,keV for the first six observations, adding a \texttt{powerlaw} component to sp1--sp6 yields a powerlaw normalization with a lower end of the 90\% confidence interval that is consistent with zero.}. 
We allowed the disk temperature and disk normalization to be different across the seven epochs. We hereafter refer to this best-fit fiducial model ($\chi^2/dof=265.13/115=2.31$) as model (A), which provides a reference point for model comparison. The fitting result is shown in the top two panels of Figure~\ref{fig:nicer_drop1_sp}.

Next, we applied the absorption model of \texttt{tbabs*zashift*xstar*continuum}. 
Following model (A), we chose \texttt{continuum=diskbb} for sp1--sp6, and \texttt{continuum=(diskbb+powerlaw)} for sp7. 
We tied log$\xi$ and $z$ in \texttt{xstar} to be the same and allowed $N_{\rm H}$ to vary across observations.
The best fit, named as model (B), gives $\chi^2/dof=151.01/106=1.42$ and represents a significant improvement over model (A). 
However, the best-fit $v_{\rm out}=-0.64\pm0.01c$, which is faster than the velocity of any ultra-fast outflows reported in the X-ray literature ($\sim$0.03--$0.59c$, \citealt{Tombesi2010, Chartas2021}), and the corresponding model (b) does not provide an acceptable fit in \S\ref{subsubsec:nicer_1keV_line}.

\begin{deluxetable*}{cc|ccccccc}[htbp!]
\tablecaption{Model (C): best-fit parameters for seven \nicer observation at $\delta t \sim\!196$\,d. \label{tab:nicer_drop1_modC1} }
\tabletypesize{\footnotesize}
\tablehead{
\colhead{Component}
& \colhead{Parameter}
& \colhead{sp1}
& \colhead{sp2}
& \colhead{sp3}
& \colhead{sp4}
& \colhead{sp5}
& \colhead{sp6}
& \colhead{sp7}
}
\startdata
\multirow{3}{*}{\texttt{xstar}} & $N_{\rm H}$ ($10^{21}\,{\rm cm^{-2}}$) 
& $0.99_{-0.14}^{+0.41}$
& $1.28_{-0.40}^{+0.62}$
& $3.56_{-1.07}^{+1.42}$
& $1.45_{-0.19}^{+0.24}$
& $1.33_{-0.15}^{+0.16}$
& $1.36_{-0.15}^{+0.18}$
& $0.37_{-0.14}^{+0.25}$
\\
& log$\xi$ (erg\,cm\,s$^{-1}$) 
& \multicolumn{7}{c}{$3.00_{-0.01}^{+0.06}$}\\
& $v_{\rm out}/c$ 
& \multicolumn{7}{c}{$-0.110_{-0.010}^{+0.010}$}\\
\hline
\multirow{2}{*}{\texttt{diskbb}} & $T_{\rm in}$ (eV) 
& $128.7_{-2.3}^{+0.9}$
& $120.8_{-1.4}^{+1.2}$
& $103.5_{-1.3}^{+1.3}$
& $102.3_{-0.9}^{+0.9}$
& $113.5_{-0.7}^{+0.8}$
& $117.9_{-0.7}^{+0.8}$
& $145.1_{-2.0}^{+0.9}$ \\
& norm$_{\rm dbb}$ ($10^3$)
& $16.74_{-0.06}^{+0.11}$
& $16.39_{-0.12}^{+0.13}$
& $20.97_{-0.16}^{+0.18}$
& $13.61_{-0.06}^{+0.05}$
& $16.02_{-0.05}^{+0.04}$
& $17.75_{-0.06}^{+0.11}$
& $11.94_{-0.04}^{+0.04}$ \\
\hline
\multirow{2}{*}{\texttt{powerlaw}} & $\Gamma$ &... &...&...&...&...&...& $3.66_{-0.24}^{+0.10}$ \\
& norm$_{\rm PL}$ ($10^{-5}$) & ...&...&...&...&...&... & 
$ 97.0_{-9.5}^{+9.6}$\\
\hline 
\texttt{gsmooth}  & $\sigma$ (keV) & \multicolumn{7}{c}{$0.066_{-0.008}^{+0.009}$}\\
\hline 
\multirow{2}{*}{\texttt{emi}} & $v_{\rm out}/c$
& \multicolumn{7}{c}{$-0.155_{-0.007}^{+0.007}$}\\
& norm$_{\rm emi}$ ($10^{-3}$) 
& $25.42_{-2.88}^{+2.32}$
& $5.25_{-1.44}^{+3.82}$
& $1.17_{-0.14}^{+0.23}$
& $0.00_{\tablenotemark{\scriptsize a}}^{0.08}$
& $0.00_{\tablenotemark{\scriptsize a}}^{0.16}$
& $0.00_{\tablenotemark{\scriptsize a}}^{0.17}$
& $18.16_{-3.42}^{+5.91}$\\
\hline 
& $\chi^2/dof$ & \multicolumn{7}{c}{102.83/97}\\
\enddata 
\tablecomments{Model (C): \texttt{tbabs*zashift*(xstar*continuum+gsmooth*emi)}, where \texttt{continuum=diskbb} for sp1--sp6, and \texttt{continuum=(diskbb+powerlaw)} for sp7.}
\tablenotetext{a}{Parameter pegged at the minimum value of the model.}
\end{deluxetable*}

Following that, we added emission from the ionized absorber by using \texttt{tbabs*zashift*(xstar*continuum+gsmooth*emi)}. We tied $\sigma$ in \texttt{gsmooth} and $z$ in \texttt{emi} to be the same and allowed norm$_{\rm emi}$ to vary across observations.
The best fit, named as model (C), has $\chi^2/dof=102.83/97=1.06$ and is shown in the middle two rows of Figure~\ref{fig:nicer_drop1_sp}. 
The best-fit parameters are presented in Table~\ref{tab:nicer_drop1_modC1}. We see that the ionization state (log$\xi$), $v_{\rm out}$ of the absorption and emission components, and $\sigma$ in the \texttt{gsmooth} component are all similar to values found in model (c) (see \S\ref{subsubsec:nicer_1keV_line}).
We note that the main resonant emission and absorption features in the soft X-ray band for an ionization parameter ${\rm log}\xi\approx\!3$ would come from O VIII, Ne IX/Ne X and Fe L. Given the high number of lines from this range of ionized species and the limited energy resolution of the detectors, the features would appear blended together.

\begin{deluxetable*}{cc|ccccccc}[htbp!]
\tablecaption{Model (D): best-fit parameters for seven \nicer observation at $\delta t \sim\!196$\,d. \label{tab:nicer_drop1_modD} }
\tabletypesize{\footnotesize}
\tablehead{
\colhead{Component}
& \colhead{Parameter}
& \colhead{sp1}
& \colhead{sp2}
& \colhead{sp3}
& \colhead{sp4}
& \colhead{sp5}
& \colhead{sp6}
& \colhead{sp7}
}
\startdata
\multirow{2}{*}{\texttt{bbody}} & $kT_{\rm bb}$ (eV) 
& $79.9_{-2.7}^{+1.8}$ 
& $79.9_{-2.2}^{+1.9}$   
& $74.2_{-1.6}^{+1.6}$  
& $76.4_{-1.1}^{+1.1}$  
& $80.9_{-0.9}^{+0.9}$  
& $83.6_{-1.2}^{+1.2}$   
& $92.2_{-5.4}^{+2.8}$  \\
& norm$_{\rm bb}$ ($10^{-5}$) 
& $57.6_{-3.6}^{+2.0}$
& $48.8_{-1.9}^{+1.4}$  
& $31.5_{-1.0}^{+1.1}$ 
& $21.5_{-0.4}^{+0.4}$ 
& $39.4_{-0.6}^{+0.6}$  
& $51.0_{-0.8}^{+0.8}$ 
& $58.4_{-7.2}^{+3.0}$\\ 
\hline 
\multirow{2}{*}{\texttt{powerlaw}} & $\Gamma$ &... &...&...&...&...&...& $3.73_{-0.08}^{+0.11}$ \\
& norm$_{\rm PL}$ ($10^{-5}$) & ...&...&...&...&...&... & 
$ 192.1_{-12.2}^{+9.9}$\\
\hline
\texttt{gsmooth}  & $\sigma$ (keV) & \multicolumn{7}{c}{$0.142_{-0.007}^{+0.009}$}\\ 
\hline 
\multirow{6}{*}{\texttt{xillverTDE}} & log$\xi$ (erg\,cm\,s$^{-1}$) & \multicolumn{7}{c}{ $2.14_{-0.02}^{+0.33}$ }\\ 
& log$n$ (cm$^{-3}$) & \multicolumn{7}{c}{ $19.00_{-0.91}^{ \tablenotemark{\scriptsize a}}$ }\\ 
& $v_{\rm out}/c$
& \multicolumn{7}{c}{$-0.340_{-0.011}^{+0.013}$}  \\
& $i$ (deg) & \multicolumn{7}{c}{45 (frozen)}\\
& $A_{\rm Fe}$ & \multicolumn{7}{c}{1 (frozen)}\\
& norm$_{\rm refl}$ ($10^{-13}$ )
& $21.8_{-0.7}^{+38.7}$ 
& $9.3_{-1.2}^{+20.0}$  
& $2.9_{-0.4}^{+8.4}$ 
& $0.5_{-0.2}^{+1.6}$  
& $2.2_{-0.3}^{+6.3}$  
& $3.0_{-0.6}^{+8.3}$ 
& $19.6_{-2.4}^{+54.3}$ \\ 
\hline 
& $\chi^2/dof$ & \multicolumn{7}{c}{111.20/104}\\
\enddata 
\tablecomments{Model (D): \texttt{tbabs*zashift*(gsmooth*xillverTDE+continuum)}, where \texttt{continuum=bbody} for sp1--sp6, and \texttt{continuum=(bbody+powerlaw)} for sp7. }
\tablenotetext{a}{Parameter pegged at the maximum value of the model.}
\tablenotetext{b}{Parameter pegged at the minimum value of the model.}
\end{deluxetable*}

Finally, we applied the reflection model of \texttt{tbabs*zashift*(gsmooth*xillverTDE+continuum)}, where \texttt{continuum=bbody} for sp1--sp6, and \texttt{continuum=(bbody+powerlaw)} for sp7. 
We allowed norm$_{\rm refl}$ to be different across observations, and tied the other three free parameters in \texttt{xillverTDE} ($\xi$, $n$, and $z$) to be the same. 
The best fit, with $\chi^2/dof =111.20/104=1.07$, is named as model (D) and shown in the bottom two rows of Figure~\ref{fig:nicer_drop1_sp}. The best-fit parameters are presented in Table~\ref{tab:nicer_drop1_modD}. We see that the best-fit log$n$, log$\xi$, and $v_{\rm out}$ are slightly different from the best-fit values of model (d) shown in \S\ref{subsubsec:nicer_1keV_line}.

\begin{figure}[htbp!]
\centering
\includegraphics[width=0.44\columnwidth]{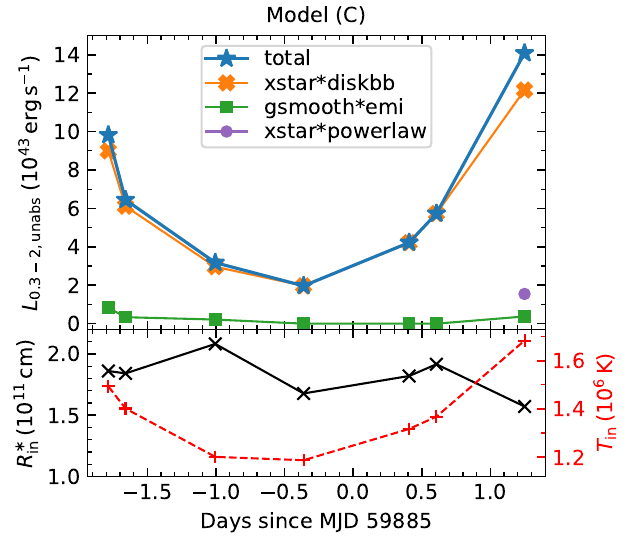}
\includegraphics[width=0.44\columnwidth]{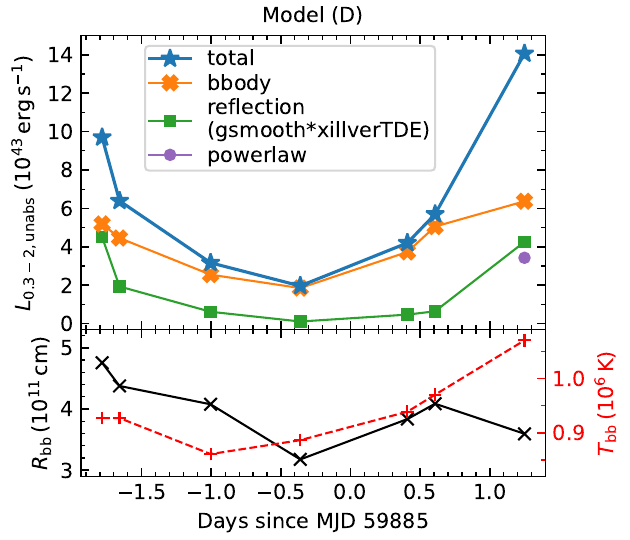}
\caption{\textit{Upper}: Evolution of 0.3--2\,keV X-ray luminosities (corrected for Galactic absorption) in models (C) and (D), shown both in total and contributions in different spectral components. 
\textit{Lower}: Evolution of temperature (``+'' connected by dashed lines) and radius (``$\times$'' connected by solid lines) evolution in the \texttt{diskbb} component in model (C) and the \texttt{bbody} component in model (D). 
\label{fig:Lx_comp}}
\end{figure}

The upper panels of Figure~\ref{fig:Lx_comp} illustrate the flux variation across the seven observations, highlighting contributions from various spectral components. 
In both models, the total luminosity (shown by the blue asterisks) initially decreased by a factor of $\approx\!5$ from sp1 to sp4, and then increased by a factor of $\approx\!7$ from sp4 to sp7. 
In model (C), the fluctuations in total luminosity mainly stems from changes in the \texttt{diskbb} component.
We note that since the $N_{\rm H}$ of the \texttt{xstar} component is relatively small (between $3.7\times 10^{20}\,{\rm cm^{-2}}$ and $3.6\times 10^{21}\,{\rm cm^{-2}}$, Table~\ref{tab:nicer_drop1_modC1}), the 0.3--2\,keV \texttt{diskbb} luminosity before absorption by the ionized absorber is only a factor of 1.02--1.18 greater than the orange crosses shown in the upper left panel of Figure~\ref{fig:Lx_comp}. 
In model (D), the variability come from variations in both the reflection component and the blackbody component. 

The lower panels of Figure~\ref{fig:Lx_comp} show the temperature and radius evolution in the thermal components in each model. 
In model (C), the inner disk temperature varies between $1.19\times 10^6$\,K and $1.68\times 10^6$\,K, and the apparent inner disk radius experiences minor variations within the range of (1.57--$2.08)\times10^{11}$\,cm. 
Fitting a function of the type log$(L_{\rm diskbb, unabs0}/{\rm erg\,s^{-1}}) = a + b \times {\rm log}(T_{\rm in}/{\rm K}) + \epsilon$ yields a slope of $b = 4.23_{-0.52}^{+0.33}$ and an intrinsic scatter of $\epsilon=0.09$, where $L_{\rm diskbb, unabs0}$ is the 0.3--2\,keV \texttt{diskbb} luminosity corrected for both Galactic absorption and absorption from the ionized absorber. 
In model (D), the blackbody temperature $T_{\rm bb}$ remains relatively constant at (0.87--$1.07)\times10^6$\,K, and the blackbody radius (inferred from the norm$_{\rm bb}$ parameter) changes within (3.2--$4.8) \times 10^{11}$\,cm. 
Fitting a function of the type log$(L_{\rm bbody, unabs}/{\rm erg\,s^{-1}}) = a + b \times {\rm log}(T_{\rm bb}/{\rm K}) + \epsilon$ yields a slope of $b = 4.09_{-0.62}^{+0.49}$ and an intrinsic scatter of $\epsilon=0.13$, where $L_{\rm bbody, unabs}$ is the 0.3--2\,keV \texttt{bbody} luminosity corrected for Galactic absorption.

To summarize, both model (C) and model (D) provide good and statistically comparable fits to the observations. A few sanity checks were performed on these models (see details in Appendix~\ref{subsec:sanity_check}). We further discuss the implications of the fitting results in \S\ref{subsubsec:discuss_spec_mods}. 
 
\section{Discussion} \label{sec:discuss}

\subsection{Super-Eddington Accretion onto a Low-mass Massive Black Hole} \label{subsec:discuss_superEdd}

\subsubsection{Basic considerations: bolometric luminosity, inclination, and long-term evolution} \label{subsubsec:basic_consideration}

The black hole mass of $M_{\rm BH}\approx 10^{5}\,M_\odot$ (\S\ref{subsec:Mbh}) is at the low end of ZTF-selected TDEs \citep{Yao2023}. The gravitational radius is $r_{\rm g}=GM_{\rm BH}/c^2 \approx 1.5\times 10^{10}\,{\rm cm}$ and the tidal radius is $R_{\rm T} \approx 3\times 10^{12}\,{\rm cm}$ (for a Sun-like star). 
The fall-back timescale is relatively short ($t_{\rm fb}\approx 13$\,d). 
The inferred effective inner disk radius $R^{\ast}_{\rm in}\approx\! 2\times 10^{11}\,{\rm cm}\sim\! 13 R_g$ from both basic phenomenological spectral modeling (Figure~\ref{fig:multiwave_evol}) and physically motivated spectral modeling (Figure~\ref{fig:Lx_comp}). As we will show later, the inclination should be very small (cos$i\sim 1$), so $R_{\rm in}$ ($\approx R_{\rm in}^{\ast}$) is close to the innermost stable circular orbit $r_{\rm ISCO}$. 
During the monitoring campaign, the photosphere of the UV and optical component receded from $\sim\!67R_{\rm T}$ to $\sim\!20R_{\rm T}$.

\begin{figure}[htbp!]
\centering
\includegraphics[width=0.75\columnwidth]{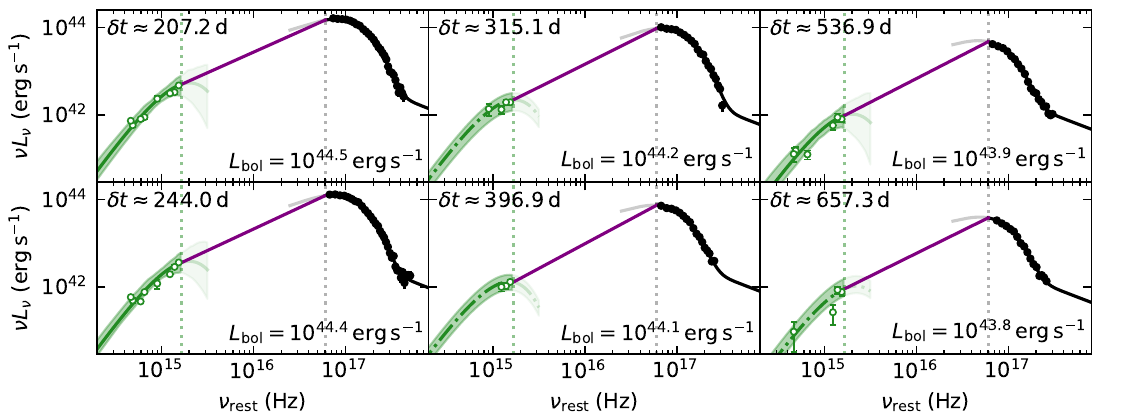}
\caption{SEDs of AT2022lri at six representative epochs (marked as dashed orange line in Figure~\ref{fig:Lbol}). The data have been corrected for extinction (in UV/optical) and Galactic absorption (in the X-ray). The green lines show the blackbody fits to the UV/optical data. Solid and dash-dotted green lines represent epochs with the blackbody temperature is fitted and fixed, respectively (see \S\ref{subsec:bbfit}). The black lines show the \nicer best-fit model with $\Gamma$ fixed at 2.8 (\S\ref{subsubsec:nicer_spec}). The purple lines are simple power-laws used as a proxy for the SED shape in the EUV band. \label{fig:SEDs} }
\end{figure}

To assess the bolometric luminosity ($L_{\rm bol}$) across the X-ray ``envelope'' (where no strong dips are observed), we selected six epochs and integrated $\int L_\nu d\nu$ from 10000\,\AA\ to 10\,keV. From 10000\,\AA\ to 1800\,\AA, we integrate below the blackbody model fitted to the UV/optical photometry (\S\ref{subsec:bbfit}). From 0.25\,keV to 10\,keV, we integrate below the best-fit model fitted to the \nicer data (\S\ref{subsubsec:nicer_spec}). From 1800\,\AA\ to 0.25\,keV, we assume that the TDE spectrum is continuous and can be approximated by a power-law of $L_\nu \propto \nu^{\alpha}$, where $\alpha$ is determined by connecting the UV/optical blackbody at 1800\,\AA\ and the X-ray model at 0.25\,keV (see Figure~\ref{fig:SEDs}). 
It is interesting to notice that the purple lines have a slope that is steeper than $\nu L_\nu \propto \nu^{4/3}$ (i.e., the standard multi-color blackbody for a constant accretion rate at all radii). A physical explanation could be that the accretion rate is not constant with radius. 
Since TDE fallback stream deposits mass near $R_{\rm T}$, the accretion rate at $R \gg R_{\rm T}$ must be much smaller than $\dot M_{\rm acc}$ at $R \leq R_{\rm T}$. This causes the emission from larger radii (at lower frequency) to be weaker than the prediction from the standard case.

\begin{figure}[htbp!]
\centering
\includegraphics[width=\columnwidth]{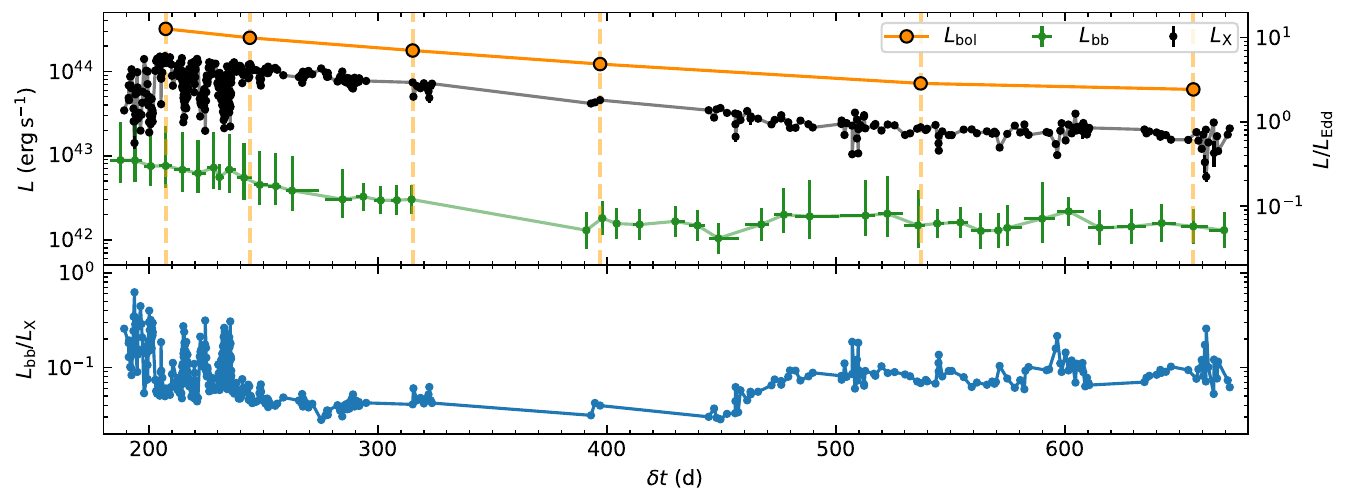}
\caption{\textit{Upper}: Evolution of bolometric luminosity $L_{\rm bol}$, 0.3--2\,keV unabsorbed X-ray luminosity $L_{\rm X}$, and blackbody luminosity of the UV/optical emission $L_{\rm bb}$. 
\textit{Lower}: Evolution of $L_{\rm bb}/L_{\rm X}$.
\label{fig:Lbol}}
\end{figure}

The upper panel of Figure~\ref{fig:Lbol} shows the evolution of $L_{\rm bol}$, which is also compared with $L_{\rm X}$ and $L_{\rm bb}$. With $M_{\rm BH}\sim\! 10^5\,M_\odot$, the bolometric luminosity declined from $3.2\times 10^{44}\,{\rm erg\,s^{-1}}\approx \!25L_{\rm Edd}$ at $\delta t\approx 207.2$\,d to $6.1\times 10^{43}\,{\rm erg\,s^{-1}}\approx \!5L_{\rm Edd}$ at $\delta t\approx 657.3$\,d. Even with the largest $M_{\rm BH}$ estimate of $\sim10^6\,M_\odot$, $L_{\rm bol}>L_{\rm Edd}$ at $\delta t<240$\,d, suggesting that AT2022lri stayed in the super-Eddington accretion regime at least during the intense rapid X-ray dipping phase. 

Recent three-dimensional (3D) general relativistic radiation magneto-hydrodynamics (GRRMHD) simulations for the TDE super-Eddington accretion flow show that copious amount of X-ray emission can only escape from the optically thin funnel along the disk axis when viewed close to face-on \citep{Dai2018, Curd2019}. \citet{Thomsen2022} performed three simulations for mass accretion rates at 7, 24, and 24\,$\dot M_{\rm Edd}$, showing that $L_{\rm bb}/L_{\rm X}\lesssim\!0.1$ can only be produced at their smallest inclination angle ($i=10^{\circ}$)\footnote{The simulations results at $i=10^{\circ}$ show even smaller values of $L_{\rm bb}/L_{\rm X}\sim\!10^{-3}$ (see Fig.~4 of \citealt{Thomsen2022}). However, this $L_{\rm bb}$ is probably underestimated due to the fact that the post-processing radiative transfer was performed in 1D, and that the inject spectrum was a single temperature $10^6$\,K blackbody, instead of a multi-temperature disk with $\sim\! 10^6$\,K inner temperature.}.

The lower panel of Figure~\ref{fig:Lbol} shows the ratio of $L_{\rm bb}/L_{\rm X}$. At $\delta t\lesssim240$\,d, it stays at $\lesssim\!0.1$ (except during the X-ray dips), pointing to a low inclination angle. 
From $\approx\!240$\,d to $\approx\!330$\,d, $L_{\rm bb}/L_{\rm X}$ exhibits a decaying trend, suggesting that the decline in UV and optical emission is faster than that in the X-ray emission. This is consistent with the expectation where as $\dot M_{\rm acc}$ decreases, the optically thin funnel (along the disk axis) gets wider, and a smaller fraction of X-ray photons are being reprocessed in the optically thick outflow \citep{Thomsen2022}.
From $\approx\!390$\,d to $\approx\!672$\,d, $L_{\rm bb}$ stays relatively constant. Similar late-time UV and optical plateaus have been observed in many other TDEs \citep{Mummery2024_late_plateau}. It is possible that at such late times, reprocessing becomes very weak such that the UV and optical emission is dominated by the outer edge of an accretion disk with a nearly constant outer radius. 

The X-ray light curve of AT2022lri is extremely variable on the timescale of $t_{\rm var}\sim 1$\,hr--1\,d during certain time intervals, such as $\delta t \sim\!190$--240\,d (Figure~\ref{fig:nicer_early_lc}), and during the XMM\,E3 observation at $\delta t \sim601$\,d (Figure~\ref{fig:xmm_lc}, right panel). However, the occurrence of such variability is episodic, as there are times where the light curve closely follows a power-law decline, as shown in Figure~\ref{fig:xray_lc} and the XMM\,E2 light curve (Figure~\ref{fig:xmm_lc}, left panel). 


At the inner disk $R_{\rm in}\sim 2\times 10^{11}$\,cm, the dynamical timescale is $t_{\rm dyn}\sim \sqrt{R_{\rm in}^3 / (GM_{\rm BH})} \sim 17$\,s, the thermal timescale is $t_{\rm th}=\frac{1}{\alpha}t_{\rm dyn}\sim 170$\,s (assuming a Shakura--Sunyaev viscosity parameter of $\alpha=0.1$), and the viscous timescale is $t_{\rm vis}=\frac{1}{\alpha}t_{\rm dyn}(H/R)^{-2}\sim 2$\,ks (assuming a disk height-to-radius ratio of $H/R=0.3$). 
The orbital timescale at $R_{\rm in}$ is $\sim\!110$\,s and at $R_{\rm T}$ is $\sim\!3$\,hr.
We see that $t_{\rm vis}$, a time in which angular momentum distribution changes due to torque caused by dissipative stresses, is comparable to the shortest observed variability timescale.

\subsubsection{Origin of the X-ray dips}\label{subsubsec:Xray_variabiltiy}

First, we rule out the idea that the X-ray variability is caused by changes in absorption along the line of sight, which is a leading model for a subtype of changing-look AGN (CLAGN) termed as changing-obscuration AGN \citep{Ricci2023}. As is shown in our spectral analysis, there is no evidence for elevated neutral intervening gas at the host redshift. In the spectral modeling with ionized absorption, the best-fit $N_{\rm H}$ values are too small to account for the large amplitude of flux variations. Therefore, although the outflow might exhibit some degree of inhomogeneities/turbulence, changing obscuration itself is not the primary reason for the strong variability.

Next, we consider if the variability can be caused by global precession, which is possible when the spin of a rotating black hole is misaligned with the angular momentum of the newly formed accretion disk, exerting a Lense–Thirring (LT) torque on the disk. Using a slim disk model for the disk structure (appropriate for the super-Eddington phase, \citealt{Strubbe2009}), \citet{Franchini2016} computed the expected precession time $T_{\rm prec}$ as functions of $M_{\rm BH}$ and $a$ (the dimensionless black hole spin parameter). For a $10^5\,M_\odot$ BH, $T_{\rm prec}$ reaches the minimum value of  3--4 days at the maximum $a$, which is much longer than the observed $t_{\rm var}$. 
Furthermore, since precession--induced light curve modulation is expected to be be periodic or quasi-periodic \citep{Stone2012}, we disfavor such a scenario. 

If the tidal disruption resides in a massive black hole binary (MBHB), the existence of a secondary MBH would cause an observable effect to the TDE fallback rates if the debris at apocenter leaves the Roche lobe of the disrupting BH. This imposes a short binary separation at sub-pc and a binary orbital period of $T_{\rm orb}\sim\!0.3$\,yr \citep{Coughlin2019_binary}. Numerical \citep{Liu2009, Ricarte2016} and hydrodynamic \citep{Coughlin2017, Vigneron2018} studies suggest that light curve dips (or interruptions) on top of a power-law decay might be expected with variability on the timescale of months-to-years. Since this is much longer than the $t_{\rm var}$ observed in AT2022lri, we disfavor a MBHB.


\begin{figure}[htbp!]
\centering
\includegraphics[width=0.7\columnwidth]{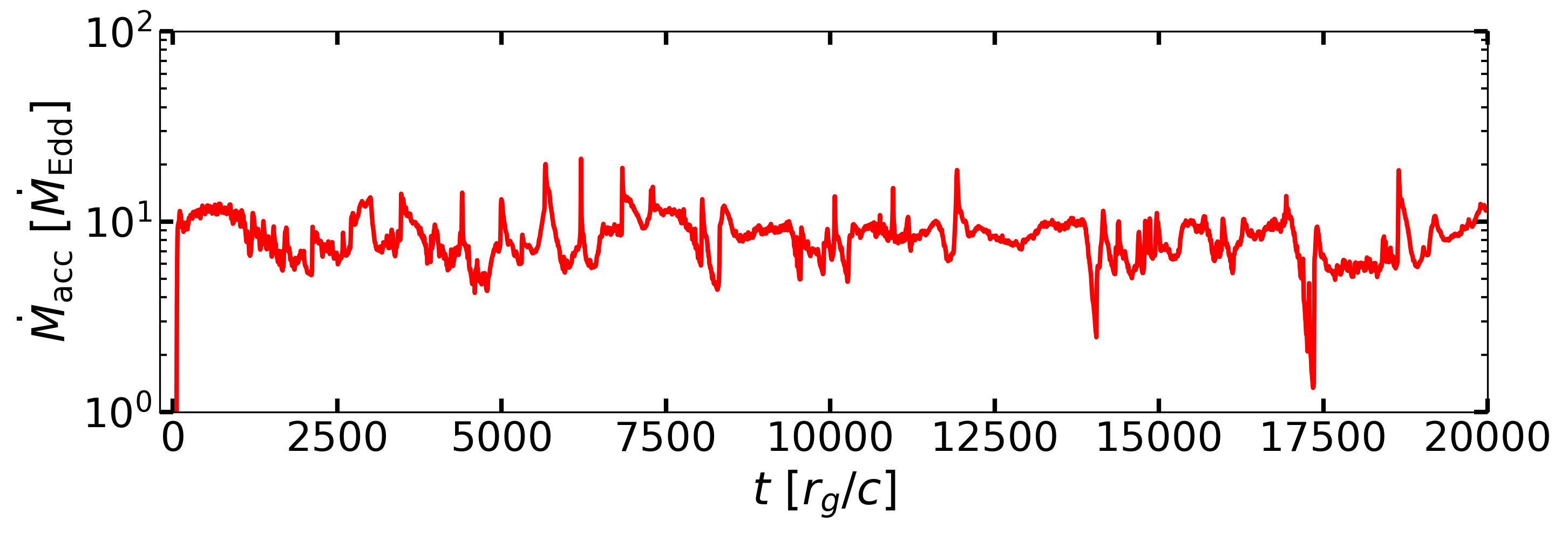}
\caption{
The mass accretion rate as a function of time in a simulated super-Eddington disk around a black hole with $M_{\rm BH}=10^6 M_\odot$ and $a=0$. The time-averaged accretion rate is around $7 \dot{M}_{\rm Edd}$. The x-axis is in unit of gravitational time $r_g /c$.
}
\label{fig:MADMdot}
\end{figure}

A natural reason for the observed fast X-ray dips is associated with episodic drops of mass accretion rates in the super-Eddington accretion flow. 
This is hinted by the fact that variations in X-ray luminosity is intrinsic to the continuum and is positively corrected with changes in $T_{\rm in}$, as seen in both basic spectral modeling (\S\ref{subsubsec:nicer_spec} and Figure~\ref{fig:nicer_xpars_corr}) and physically motivated modeling results (\S\ref{subsubsec:dip} and Figure~\ref{fig:Lx_comp}).
The decrease in $T_{\rm in}$ reflects instantaneous changes of local heating rate and mass accretion rate at $R_{\rm in}$, which might be triggered by instabilities in the magnetic field structure. 
For example, in magnetically arrested disks (MADs; \citealt{Igumenshchev2003, Narayan2003, Igumenshchev2008}), large-scale magnetic field lines threading the inner accretion flow grow to a saturation value of the (dimensionless) magnetic flux $\phi_{\rm BH}$ near the black hole horizon. Then the gas accretion process is highly governed by the interplay between magnetic flux and gas. 
Numerical simulations show in MADs  $\phi_{\rm BH}$ exhibits substantial fluctuations around the saturated value, which therefore introduces large variability in the accretion rate $\dot{M}_{\rm acc}$ \citep{Tchekhovskoy2011, Tchekhovskoy2012, McKinney2012}. 
Recent simulations of MAD disks in the super-Eddington accretion regime demonstrate similar behaviors \citep{Sadowski2016, Dai2018, Thomsen2022, Curd2023_mad_disk}. As a demonstration, we show the simulated $\dot{M}_{\rm acc}$ from a super-Eddington accretion disk around a $10^6 M_\odot$ black hole in Figure~\ref{fig:MADMdot}. The simulation has otherwise the same set up as the lowest Eddington ratio run with $\dot{M}_{\rm acc}\sim 7\dot{M}_{\rm Edd}$ in \citet{Thomsen2022}, except that the black hole spin is set to be $a=0$. One can see that the variability over a timescale of $1000 \ r_g/c \sim 1$ hr can reach from a few to $\sim\! 10$ times, which is similar to the observed scale of variability.

Furthermore, in the simulations conducted by \citet{Curd2023_mad_disk} where $\dot M_{\rm acc}$ decreases from $3\dot M_{\rm Edd}$ to $0.3 \dot M_{\rm Edd}$, this fluctuation in $\phi_{\rm BH}$ is directly linked to variations in $L_{\rm bol}$, irrespective of the BH spin. 
However, \citet{Curd2023_mad_disk} shows that quasi-periodic variability is developed in the $a=0.9$ simulation, whereas the variability appears stochastic in the $a=0$ simulation. This suggests that AT2022lri should be a TDE happening around a low-spin black hole, which is also consistent with the fact that no evidence of on-axis relativistic jets (in the form of bright hard X-ray emission) has been observed from this system. 


Another possible cause of the X-ray drops might be related to wobbling of the inner accretion disk along the MBH's spin axis. 
Interestingly, the host galaxy's disk component has an axis-ratio close to unity ($q=0.96$, \S\ref{subsec:galfits}), suggesting that the MBH's spin axis is probably along our line of sight. If the TDE disk's initial axis does not align with the spin axis, LT precession can naturally occur.
Although we discussed previously that global precession of the entire disk only gives $t_{\rm var}\gtrsim {\rm few}$ days, mass infall on shorter timescales is expected if LT torques are strong enough to tear the wrapped disk into inner and outer parts \citep{Nixon2012}. While this effect is shown to be most pronounced at low accretion rates, thinner disks, and larger oriented inclinations \citep{Raj2021_simulation}, it is still likely to happen in a thick disk if $H/R \lesssim 0.1$. In this case, the apparent drops of $T_{\rm in}$ may come from either changes in mass accretion rate or a view-angle effect, where at larger inclination angles the disk temperature measured in the soft X-ray band appears lower \citep{Dai2018}. While quasi-periodic behavior is observed in the thin disk simulation by \citet{Raj2021_implication}, one may imagine that the TDE accretion flow starts from a more asymmetric initial condition and the inner and outer disks, instead of being discrete regions, might be weakly connected by tenuous gas, rendering a more stochastic variability. Under such a circumstance, the decrease of X-ray dipping amplitude and occurrence rate after $\delta t \sim 240$\,d (see Figure~\ref{fig:xray_lc} and Figure~\ref{fig:multiwave_evol}) might signify a time when the inner disk is aligned to the black hole spin and the unstable wobbling region moves to larger radii. 

\subsubsection{Physically motivated spectral models}  \label{subsubsec:discuss_spec_mods}
\begin{figure}[htbp!]
    \centering
    \includegraphics[width=0.49\columnwidth]{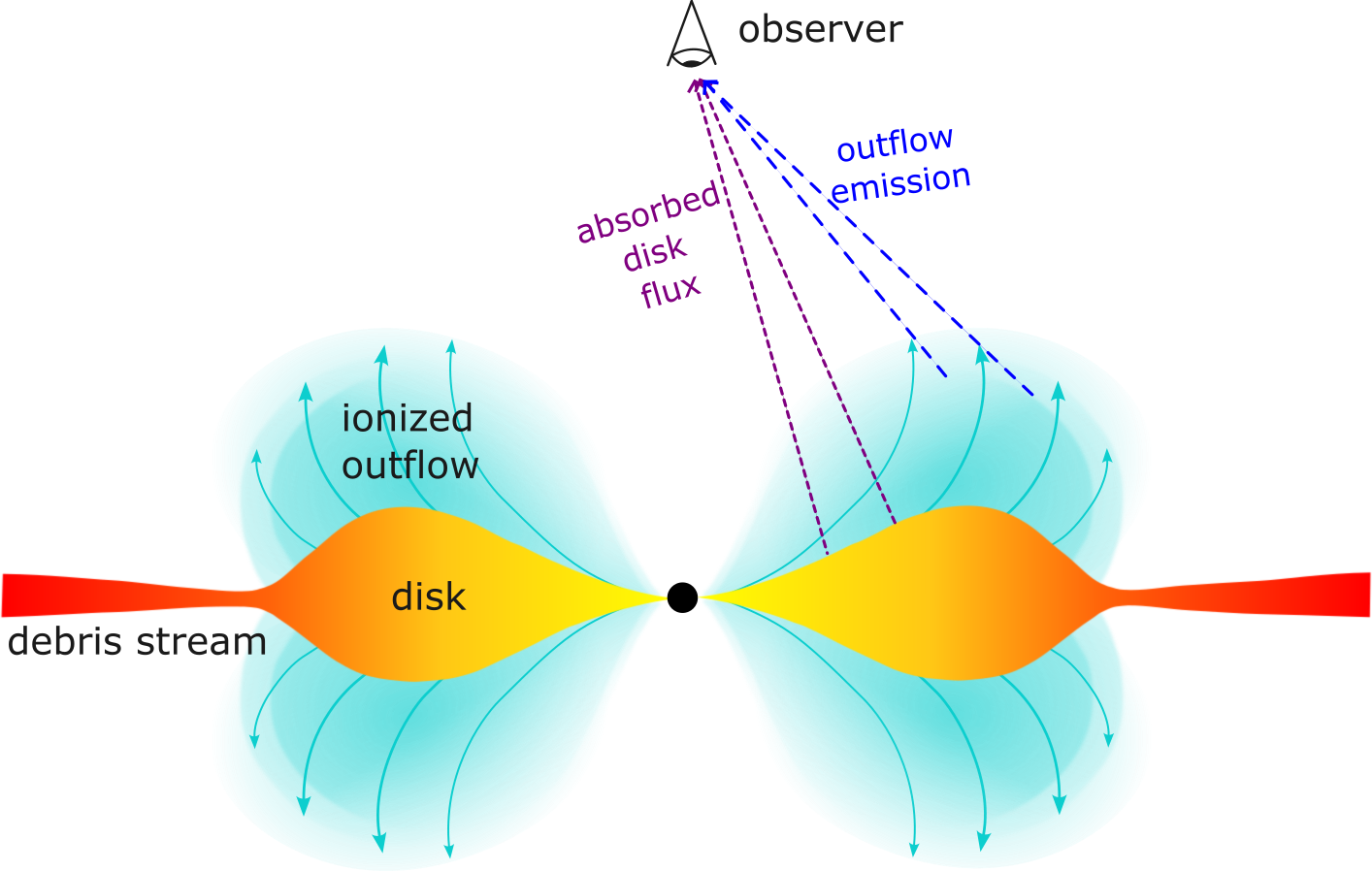}
    \includegraphics[width=0.49\columnwidth]{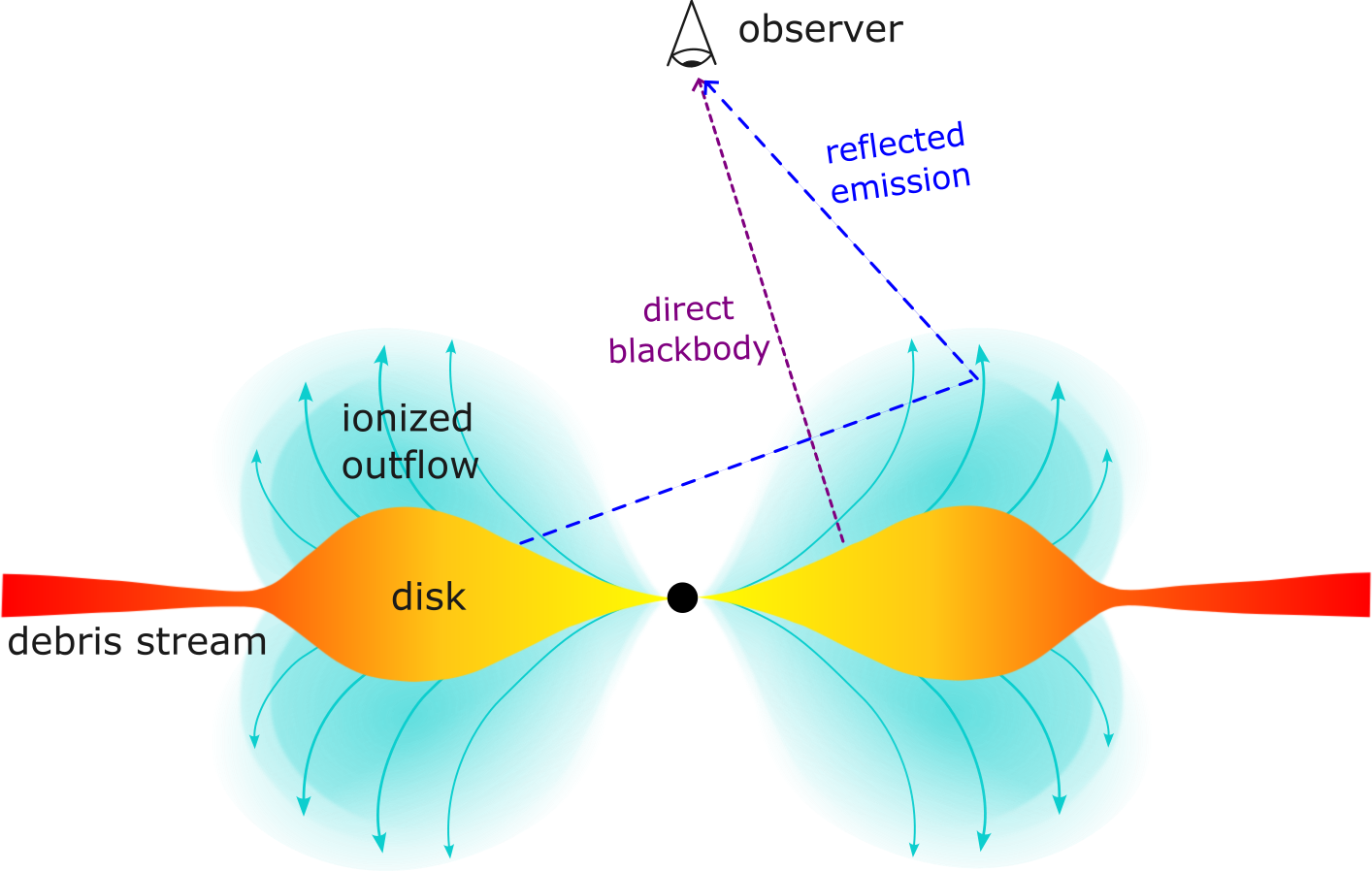}
    \caption{\textit{Left}: A schematic picture showing the geometry of the super-Eddington accretion flow, appropriate for our modeling with an ionized outflow as the absorber.
    The short dashed purple lines depict line-or-sights where the emission is mainly absorbed disk flux, and the long dashed blue lines highlight line-or-sights that primarily give rise to the emission component of the outflow.  
    \textit{Right}: A schematic picture for the reflection modeling. 
    \label{fig:cartoon}}
\end{figure}

Below 1\,keV, the spectral residual observed in our \xmm/EPIC (Figure~\ref{fig:xmm_pn_spec}) and \nicer (e.g., Figures~\ref{fig:nicer_highstat_specs}, and \ref{fig:nicer_drop1_sp}) data of AT2022lri show a flux excess around 0.5--0.6\,keV and a flux deficiency around 0.7--0.8\,keV, which look similar to the residual shapes observed in two previously known X-ray bright TDEs (ASASSN-14li and AT2020ksf) that have been modeled with absorption from blueshifted, ionized outflows \citep{Kara2018, Ajay2024, Wevers2024}. 
In AT2022lri, the wider spectral energy ranges reveal that the residuals are even stronger at 1--2\,keV. Therefore, when modeled with an ionized absorber, emission from the ionized material is further required. 
In the best-fit model for both the single \nicer observation (model (c) in \S\ref{subsubsec:nicer_1keV_line}) and the seven observations that form an X-ray dip (model (C), Table~\ref{tab:nicer_drop1_modC1}), the outflow velocity from the emission component is $v_{\rm out}\in (-0.16c, -0.13c)$, which is slightly larger than the velocity from the absorption component ($v_{\rm out}\approx -0.11c$). This can be realized for certain geometry where the outflow along the polar region has lower velocities than that along the equatorial region (see a schematic picture in the left panel of Figure~\ref{fig:cartoon}). 

Taking a fiducial outflow velocity of $v_{\rm out}\sim0.15c$, we estimate the outflow launching radius to be the distance at which the observed velocity is equivalent to the escape velocity from the BH: $R_{\rm launch}=2GM_{\rm BH}/v_{\rm out}^2\sim 1.3\times 10^{12}\,{\rm cm}\sim 90 R_{\rm g}$. The mass outflow rate can be estimated with $\dot M_{\rm out}=4\pi C_{\rm f} R_{\rm launch} N_{\rm H}\mu m_{\rm p} v_{\rm out}$ \citep{Pasham2024_QPOut}, where $C_{\rm f}$ is the covering fraction of the outflow and $\mu = 1.4$ is the mean atomic mass per proton. Taking $C_{\rm f}=0.5$, as the fiducial value from the literature, we have $\dot M_{\rm out}\sim 8.7\times 10^{19}\,{\rm g\,s^{-1}}\sim 1.4\times 10^{-6}\,M_\odot\,{\rm yr^{-1}}$. The kinetic power of the outflow $\dot E_{\rm out}=1/2 \dot M_{\rm out} v_{\rm out}^2\sim 8.9\times 10^{38}\,{\rm erg\,s^{-1}}$. 
Taking $\eta =0.1$, the mass accretion rate $\dot M_{\rm acc}=L_{\rm bol}/(\eta c^2) \sim 3.6\times 10^{24}\,{\rm g\,s^{-1}}\sim 5.7\times 10^{-2}\,M_\odot \,{\rm yr^{-1}}$. 
We see that the mass outflow rate is a tiny fraction ($\sim2.4\times 10^{-5}$) of the inflow rate. We note that the mass outflow rate estimated above should be taken as a lower limit because the launching radius could be larger than the one calculated assuming the escape velocity. Moreover, since the viewing angle is close to face-on, the column density in the absorber ($N_{\rm H}\sim 10^{21}\,{\rm cm^{-2}}$) along the line of sight will be smaller than that along other directions if the bulk of the matter ejection occurs along the equatorial plane. 

A schematic of the reflection model is shown in the right panel of Figure~\ref{fig:cartoon}. 
In the traditional reflection model \texttt{xillver}, the gas density is typically fixed at $n_e = 10^{15}\,{\rm cm^{-3}}$ \citep{Garcia2010, Garcia2013}. 
\citet{Garcia2016} considered the change of free-free emissivity, showing that high densities lead to a hotter and more ionized atmosphere, thereby strengthening the thermal continuum at $\lesssim2$\,keV. 
It was found that relativistic reflection off a high density disk could explain the soft X-ray excess observed in many Seyfert galaxies \citep{Jiang2019_highn_reflection}.
A potential issue of our modeling with \texttt{xillverTDE} is that the inferred gas density in model (D) is close to the maximum value of the model grid (i.e., $n\sim 10^{19}\,{\rm cm^{-3}}$). 
However, gas densities observed in GRRMHD simulations of super-Eddington accretion disks around TDEs are generally much lower ($n\lesssim 10^{13}\,{\rm cm^{-3}}$). Further development and extensive testing of this model may be necessary to assess how the current assumptions impact parameter constraints.

\subsection{Comparison with other TDEs and Nuclear Transients}

Here we compare AT2022lri with other MBH-powered transients with fast X-ray variability ($t_{\rm var}\lesssim 1$\,d), including jetted TDEs (\S\ref{subsec:compare_jettedTDE}), non-jetted TDEs (\S\ref{subsec:compare_nonjettedTDE}), as well as other nuclear transients with similar properties (\S\ref{subsec:compare_clagn}). 

\subsubsection{Jetted TDEs} \label{subsec:compare_jettedTDE}

\begin{figure}[htbp!]
	\centering
	\includegraphics[width=0.95\columnwidth]{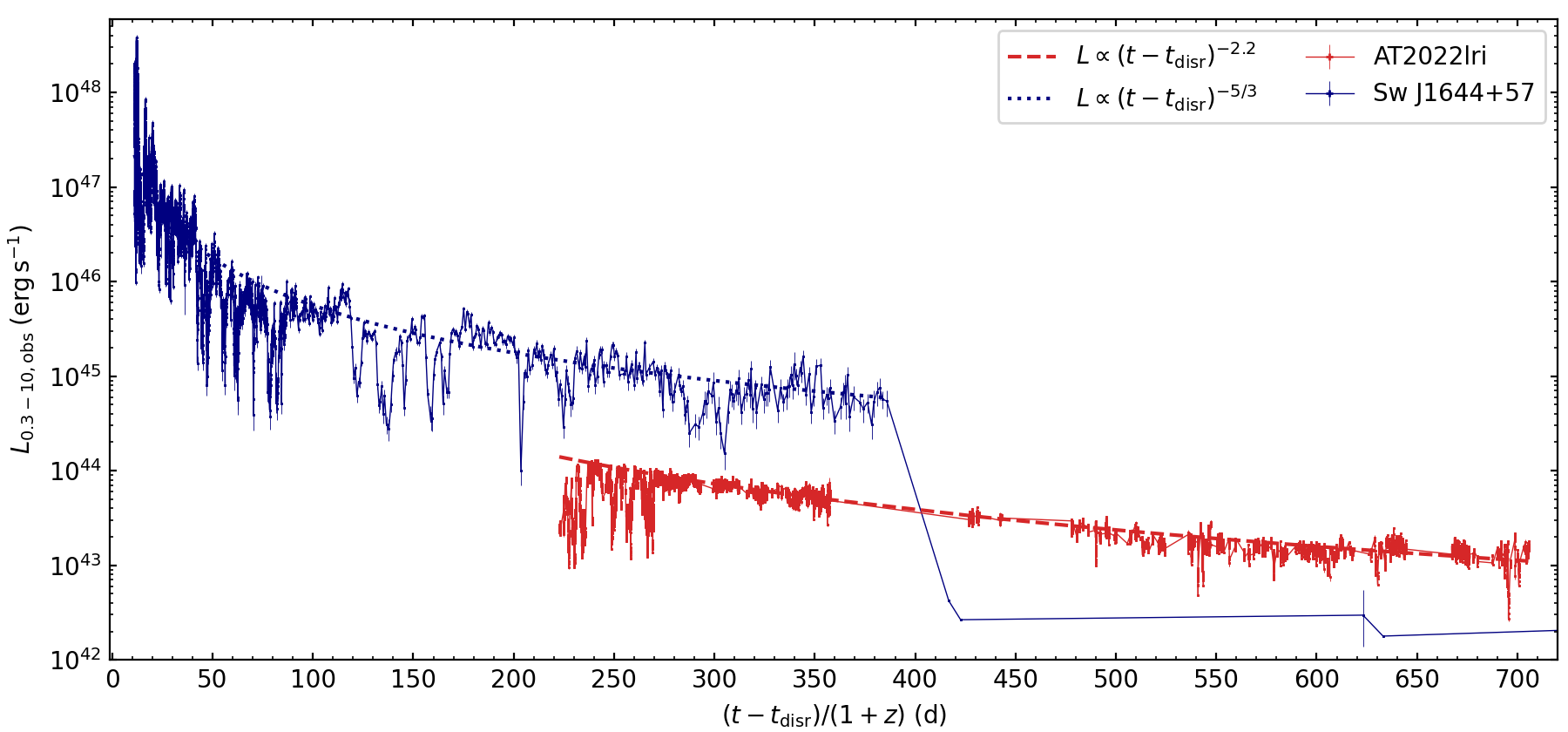}
	\caption{Isotropic equivalent X-ray luminosity of AT2022lri and Sw\,J1644+57 \citep{Mangano2016, Levan2016, Eftekhari2018} in the observer frame 0.3--10\,keV, plotted as a function of rest-frame days since the assumed disruption epoch $t_{\rm disr}$. 
	For AT2022lri we adopt $t_{\rm disr}= 59647$ (in MJD; see \S\ref{subsubsec:nicer_spec}); for Sw\,J1644+57 we adopt $t_{\rm disr} = t_{\rm trig}-15$\,d \citep{Tchekhovskoy2014}, where $t_{\rm trig}$ is 
     MJD 55648.5401.
		\label{fig:compare_J1644}}
\end{figure}

First, we compare AT2022lri with TDEs associated with on-axis relativistic jets. Among the four known jetted TDEs, sub-hour timescale X-ray variability has been observed in  Sw\,J1644+57 \citep{Burrows2011, Bloom2011, Levan2011} and AT2022cmc \citep{Pasham2023, Yao2024}. 

As shown in Figure~\ref{fig:compare_J1644}, the X-ray light curve of AT2022lri bears some resemblance to that of Sw\,J1644+57 in that both objects exhibit peak X-ray luminosities above the Eddington limit, and X-ray dips with $t_{\rm var}<1$\,hr on top of power-law light curve declines. 
Blueshifted emission line \citep{Kara2016} and hints of periodicity have been reported in Sw\,J1644+57 \citep{Saxton2012_J1644, Lei2013, Shen2014, Franchini2016}. 
\citet{Saxton2012_J1644} reported quasi-periodicity at multiple periods, which is thought to come from modulation of the jet luminosity by the disk precession and nutation.
\citet{Lei2013} argues that there exists a $T=2.7$\,d quasi-periodicity that lasts about 10\,d, which might be related to MAD around rapidly spinning massive BHs \citep{Tchekhovskoy2014} or instabilities at the nozzle point when tidally disrupted star returns to pericenter \citep{DeColle2012}. As noted by \citet{Curd2023_mad_disk}, it is possible that flux eruption events in MAD drives the X-ray variability after $\sim\!10$\,d (when a spin-aligned jet is established).

We note that despite the similarities between the X-ray light curves of AT2022lri to that of Sw\,J1644+57, the emission mechanisms are different. Jetted TDEs generally exhibit much harder X-ray spectra that can be modeled with synchrotron or inverse Compton processes, probably powered by internal energy dissipation with the jet \citep{Burrows2011, Yao2024}. The lack of hard X-ray emission in AT2022lri suggests the absence of a collimated relativistic jet. 

\subsubsection{Non-jetted TDEs} \label{subsec:compare_nonjettedTDE}

X-ray variability of TDEs is a subject of vigorous investigation. 

Large-amplitude (i.e., flux variation with a multiplicative factor $\gtrsim5$) X-ray variability on the timescale of a few days has been previously observed in the TDE 
SDSS\,J1201+30 \citep{Saxton2012_J1201}, 
2MASX\,0740-85 \citep{Saxton2017}, 
OGLE16aaa \citep{Kajava2020},
AT2019azh \citep{Hinkle2021},
AT2019ehz \citep{vanVelzen2021, Guolo2024},
AT2020ocn \citep{Pasham2024_20ocn}, 
and the TDE candidates AT2019avd \citep{Wang2023_19avd}.
They have been interpreted as TDEs in MBHBs \citep{Liu2014_J1201, Shu2020}, LT precession, brief glimpses through a patchy reprocessing layer with ``gaps'', and clumpy outflows at $\sim\!500R_{\rm g}$ from the BH in supercritical accretion \citep{Takeuchi2013, Kobayashi2018}. However, as elaborated in \S\ref{subsubsec:Xray_variabiltiy}, these explanations encounter challenges in accounting for the variability timescales and X-ray spectra observed in AT2022lri.

Moderate amplitude (i.e., flux variation with a multiplicative factor $\gtrsim2$) variability on hour timescale is seen in the TDE candidate eRASSt\,J234402.9-352640 \citep{Homan2023}. However, the lower cadence in the X-ray light curve of eRASSt\,J234402.9-352640 prohibits identification of dips atop a power-law decline, and it is not clear if the absence of spectral features is a result of the low SNR. We do note that the black hole mass of eRASSt\,J234402.9-352640 is much greater ($M_{\rm BH}\sim10^8\,M_\odot$), indicating $\lambda_{\rm Edd}< 0.1$ and perhaps a geometrically thinner disk. 


\subsubsection{Other Nuclear Transients} \label{subsec:compare_clagn}

Using similar X-ray spectral analysis techniques as adopted in our modeling with an ionized absorber, sub-relativistic outflows have been detected in a few AGN that exhibit X-ray dips on the timescale of $\sim\!$months, such as the radio-loud AGN 3C 111 \citep{Tombesi2012} and the radio-quiet AGN PG 1448+273 \citep{Laurenti2021}. The former has a black hole mass of $M_{\rm BH}\sim\!2\times 10^8\,M_\odot$ \citep{Chatterjee2011}, whereas the latter has $M_{\rm BH}\sim\!2\times 10^7\,M_\odot$. The outflows in these AGNs are of much higher column densities $N_{\rm H}\sim 10^{23}\,{\rm cm^{-2}}$ and ionization (log$\xi\sim\!5$). It has been proposed that these X-ray dips could be directly linked to a depletion of the inner accretion disk caused by the material being expelled through the outflow \citep{Tombesi2012, Laurenti2021} --- a picture that is different from what we propose in \S\ref{subsubsec:Xray_variabiltiy}.

The 1\,keV residual feature seen in AT2022lri looks reminiscent to the 2018--2019 X-ray observations of 1ES 1927+654. This object was a well-known type 2 AGN (see \citealt{Gallo2013} and references therein). On 2017 December 23, the ASASSN survey detected an optical flare, the light curve evolution of which follows the canonical slow rise and power-law decay normally observed in TDEs \citep{Trakhtenbrot2019_1ES}. Extensive follow-up X-ray observations starting from May 2018 revealed that the pre-existing AGN corona (signified by the power-law component) was first destroyed in the event and was recreated around one year after the optical discovery \citep{Ricci2020}. The host black hole has $M_{\rm BH}\sim\!10^6\,M_\odot$ \citep{Li2022_1ES}.
\citet{Ricci2021} presented the short-timescale ($t_{\rm var}<5$--10\,ks) large-amplitude X-ray variability, studied the spectral evolution, and modeled the 1\,keV spectral feature using both symmertric Gaussian lines and relativistically broadened asymmetric line profiles (i.e., the \texttt{relline} model, \citealt{Dauser2010, Dauser2013}).
It was found that the X-ray luminosity increases with the disk temperature, with an average power-law slope of $b\approx3.85$ (a relation similar to what we see in AT2022lri, see Figure~\ref{fig:nicer_xpars_corr}), though the trend is steeper (flatter) at lower (higher) luminosities in 1ES 1927+654.
\citet{Masterson2022} applied \texttt{xillverTDE} and successfully explained the spectral shape using blurred reflection, with best-fit spectral parameters similar to what we found in model (D).
1ES 1927+654, as a special changing-state AGN with no prior analogs, has been interpreted as a TDE in a preexisting AGN \citep{Trakhtenbrot2019_1ES, Ricci2020} and a ``magnetic flux inversion'' event \citep{Laha2022}. Given the similarities between AT2022lri and 1ES 1927+654, it might be the same physical mechanism that drives the fast variability in both systems.

\section{Summary and Conclusions} \label{sec:summary}
We have presented an extensive multi-wavelength study of the TDE AT2022lri. The main results from our work are:
\begin{enumerate}
    \item At a redshift of $z=0.03275$, the host of AT2022lri is a quiescent galaxy that can be modeled with a disk component viewed face-on and a pseudo bulge component.
    It has a total stellar mass of $\sim10^{9.6}\,M_\odot$, a black hole mass of $\sim 10^{5}\,M_\odot$, and no historical X-ray detection at $\lesssim 2.5\times 10^{41}\,{\rm erg\,s^{-1}}$ (\S\ref{sec:hostgalaxy}).
    \item AT2022lri belongs to the TDE-H+He optical spectroscopic subtype (\S\ref{subsec:opt_spec}). 
    \item The X-ray light curve from 187\,d to 672\,d exhibits dips atop a power-law that declines from $1.5\times 10^{44}\,{\rm erg\,s^{-1}}$ to $1.8\times 10^{44}\,{\rm erg\,s^{-1}}$. 
    The dips have amplitudes on the order of $\approx 2$--8 and timescales of $\approx 0.5$\,hr--1\,d. 
    Fast X-ray variability shows up at intermittent phases and persists throughout our monitoring program from 187\,d to 672\,d (see Figure~\ref{fig:xray_lc}).
    \item The bolometric luminosity remains above the Eddington limit at $\delta t \lesssim 240$\,d (and perhaps beyond, see Figure~\ref{fig:Lbol}). The ratio between the UV and optical blackbody luminosity to that of the X-ray luminosity remains small ($L_{\rm bb}/L_{\rm X}\lesssim 0.1$), suggesting a super-Eddington accretion flow viewed face-on. 
    This interpretation is fairly insensitive to the inferred BH mass within a reasonable range of $10^4$--$10^6\,M_\odot$. 
    \item When fitted with simple continuum models, the X-ray spectra of AT2022lri exhibit a strong residual that peaks around 1\,keV. The spectral features can be well modeled with either absorption and emission from a blueshifted ($v_{\rm out}\sim\!0.1c$) ionized absorber ($N_{\rm H}\sim10^{21}\,{\rm cm^{-2}}$, log$\xi\sim\!3$), or reflection off a dense outflow ($v_{\rm out}\sim0.3c$). These two models are statistically comparable and physically feasible.
    Both models suggest the existence of sub-relativistic outflows that are consistent with various simulations of disk winds from super-Eddington accretion disks.
    \item There is no clear evidence of narrow absorption lines similar to those observed in ASASSN-14li \citep{Miller2015} and ASASSN-20qc \citep{Kosec2023} in the RGS spectrum (\S\ref{subsubsec:rgs}). 
    \item The intermittent strong X-ray dips correspond to drops of the inner disk temperature. We propose that this is a result of episodic drops of mass accretion rates at the inner disk triggered by magnetic instability or/and wobbling of the inner disk along the BH's spin axis.
\end{enumerate}

In the future, continued observations of AT2022lri are important to further track the evolution of the accretion flow and reveal possible state transitions. 
An in-depth X-ray timing analysis is needed to search for quasi-periodic oscillations and reverberation signals, and to reveal connections to other accreting black hole systems. 
Radio observations of AT2022lri will be particularly useful to probe the galaxy circumnuclear matter density profile and the outflow velocity evolution \citep{Alexander2020}, which might serve as a diagnostic for the two spectral models. 

Looking ahead, a systematic timing analysis on all TDEs with high cadence observations is needed to address the ubiquity and physical origins of short-timescale X-ray variability in TDEs. Next-generation X-ray instruments such as AXIS will be able to extend the energy coverage and potentially directly differentiate the spectral models \citep{Reynolds2023, AXIStdamm2023}.

\vspace{1cm}

\textit{Acknowledgements} -- 
We thank Norbert Schartel for approving our XMM-Newton ToO requests. 
We thank Andrew Mummery, Chris Nixon, Nick Stone, Anil Seth, Luis C. Ho, Hua Feng, Peter Kosec, and Junjie Mao for helpful discussions. 

YY and SRK acknowledge support from NASA under awards No. 80NSSC22K1347 and No. 80NSSC24K0534.
MG and SG are supported in part by NASA XMM-Newton grants 80NSS23K06215 and 80NSSC22K0571.
FT acknowledges funding from the European Union --- Next Generation EU, PRIN/MUR 2022 (2022K9N5B4).
RL was supported by the National Science Foundation of China (11721303, 11991052, 12011540375, 12233001), the National Key R\&D Program of China (2022YFF0503401), and the China Manned Space Project (CMS-CSST-2021-A04, CMS-CSST-2021-A06). 
LD and TK acknowledge support from  the National Natural Science Foundation of China and the Hong Kong Research Grants Council (12122309, 17305920, 17314822, 27305119).
This research was supported in part by grant no. NSF PHY-2309135 to the Kavli Institute for Theoretical Physics (KITP).

This work is based on observations obtained with the Samuel Oschin Telescope 48-inch and the 60-inch Telescope at the Palomar Observatory as part of the Zwicky Transient Facility project. ZTF is supported by the National Science Foundation under Grant No. AST-2034437 and a collaboration including Caltech, IPAC, the Weizmann Institute of Science, the Oskar Klein Center at Stockholm University, the University of Maryland, Deutsches Elektronen-Synchrotron and Humboldt University, the TANGO Consortium of Taiwan, the University of Wisconsin at Milwaukee, Trinity College Dublin, Lawrence Livermore National Laboratories, IN2P3, University of Warwick, Ruhr University Bochum, Cornell University, and Northwestern University. Operations are conducted by COO, IPAC, and UW.

SED Machine is based upon work supported by the National Science Foundation under Grant No. 1106171.

The ZTF forced-photometry service was funded under the Heising-Simons Foundation grant \#12540303 (PI: Graham).

This work has made use of data from the Asteroid Terrestrial-impact Last Alert System (ATLAS) project. The Asteroid Terrestrial-impact Last Alert System (ATLAS) project is primarily funded to search for near earth asteroids through NASA grants NN12AR55G, 80NSSC18K0284, and 80NSSC18K1575; byproducts of the NEO search include images and catalogs from the survey area. This work was partially funded by Kepler/K2 grant J1944/80NSSC19K0112 and HST GO-15889, and STFC grants ST/T000198/1 and ST/S006109/1. The ATLAS science products have been made possible through the contributions of the University of Hawaii Institute for Astronomy, the Queen’s University Belfast, the Space Telescope Science Institute, the South African Astronomical Observatory, and The Millennium Institute of Astrophysics (MAS), Chile.

A major upgrade of the Kast spectrograph on the Shane 3~m telescope at Lick Observatory, led by Brad Holden, was made possible through generous gifts from the Heising-Simons Foundation, William and Marina Kast, and the University of California Observatories. Research at Lick Observatory is partially supported by a generous gift from Google.


\software{
\texttt{astropy} \citep{Astropy2013},
\texttt{emcee} \citep{Foreman-Mackey2013},
\textsc{GALFITS} (Li \& Ho, in prep),
\texttt{heasoft} \citep{HEASARC2014},
\texttt{LPipe} \citep{Perley2019lpipe}, 
\texttt{matplotlib} \citep{Hunter2007},
\texttt{SAS} \citep{Gabriel2004},
\texttt{scipy} \citep{Virtanen2020},
\texttt{xspec} \citep{Arnaud1996}.
}

\appendix
\section{Photometry and Observing Logs} \label{sec:table_log}

The UV and optical photometry of AT2022lri is given in Table~\ref{tab:phot}. Note that the UVOT photometry is host subtracted. The host galaxy UV AB magnitudes (corrected for Galactic extinction) from our SED model is $uvw2_{\rm UVOT}=22.30\pm0.05$, $uvm2_{\rm UVOT} = 22.09\pm0.05$, $uvw1_{\rm UVOT} = 20.78\pm0.05$, and $U_{\rm UVOT} = 19.03\pm0.05$.

\begin{deluxetable}{ccccc}[htbp!]
\tablecaption{UV and optical photometry of AT2022lri. \label{tab:phot}}
\tabletypesize{\footnotesize}
	\tablehead{
		\colhead{MJD} &
		\colhead{Instrument} &
		\colhead{Filter} &
		\colhead{$f_\nu$ ($\mu$Jy)} &
		\colhead{$\sigma_{f_\nu}$ ($\mu$Jy)} 
	}
	\startdata
	59874.9059 & ATLAS & $o$ & $63.75$ & $5.49$ \\
	59875.6055 & UVOT & $uvm2$ & $119.69$ & $7.77$ \\
	59875.6080 & UVOT & $uvw1$ & $108.98$ & $8.34$ \\
	59875.6094 & UVOT & $U$ & $99.99$ & $13.40$ \\
	59875.6128 & UVOT & $uvw2$ & $161.98$ & $7.77$ \\
	59875.6171 & ATLAS & $c$ & $70.24$ & $3.85$ \\
	59880.3359 & ZTF & $r$ & $75.75$ & $7.84$ \\
	59880.3567 & ZTF & $g$ & $68.67$ & $5.48$ \\
	\enddata
	\tablecomments{$f_\nu$ is flux density corrected for Galactic extinction. 
	(This table is available in its entirety in machine-readable form.)}
\end{deluxetable}

A log of low-resolution optical spectroscopic observation is given in Table~\ref{tab:spec}. 

\begin{deluxetable}{crccccr}[htbp!]
\tablecaption{Log of AT2022lri low-resolution optical spectroscopy. \label{tab:spec}}
\tabletypesize{\footnotesize}
	\tablehead{
		\colhead{Start Date}  
		& \colhead{$\delta t$ (days)}
		& \colhead{Telescope}
		& \colhead{Instrument}
		& \colhead{Wavelength range (\AA)}
		& \colhead{Slit width ($^{\prime\prime}$)}
		& \colhead{Exp. (s)} 
	}
 \startdata
2022-09-05.3 & 140.7 & P60 & SEDM & 3770--9223 & -- & 2250\\
2022-09-15.3 & 150.4 & P60 & SEDM & 3770--9223 & -- & 2250 \\
2022-09-26.3 & 161.0 & P60 & SEDM & 3770--9223 & -- & 2250 \\
2022-10-11.4 & 175.6 & P200 & DBSP & 3200--5550, 5750--9995 & 1.0 & 1800 \\
2022-10-26.4 & 190.2 & P200 & DSBP & 3200--5550, 5750--9995 & 1.5 & 1200 \\
2022-10-31.3 & 194.9 & LDT  & De Veny & 3586--8034 & 1.5 & 2800 \\ 
2022-11-16.3 & 210.4 & Lick-3m & Kast & 3525--10500 & 2.0 & 3660/3600\tablenotemark{a} \\
2022-11-17.4 & 211.5 & Keck-I & LRIS & 3200--10250 & 1.5 & 600 \\ 
2022-11-18.2 & 212.3 & P200 & DBSP & 3200--5550, 5750--9995 & 1.5 & 1500 \\
2022-11-30.3 & 224.0 & Lick-3m & Kast & 3525--10500 & 2.0 & 3660/3600\tablenotemark{a} \\
2023-01-16.3 & 269.5 & Keck-I & LRIS & 3200--10250 & 1.0 & 600\\
2023-08-25.5 & 483.7 & Lick-3m & Kast & 3525--10500 & 2.0 & 3660/3600\tablenotemark{a} \\
2023-10-17.5 & 535.0 & Keck-I & LRIS & 3200--10250 & 1.0 & 900\\
 \enddata 
\tablecomments{All spectra will be made available on the TNS page of this source (\url{https://www.wis-tns.org/object/2022lri}) at the time of manuscript acception. }
\tablenotetext{a}{Exposure times on blue/red sides of the spectrograph.}
\end{deluxetable}

A log of \xmm observations is given in Table~\ref{tab:xmm-log}

\begin{deluxetable}{ccccc}[htbp!]
\tablecaption{\xmm Observation Log \label{tab:xmm-log}}
\tabletypesize{\footnotesize}
	\tablehead{
		\colhead{Index}
		& \colhead{obsID}  
		& \colhead{Exp.} 
		& \colhead{Start Date} 
		& \colhead{$\delta t$}\\
		\colhead{}
		& \colhead{}
		& \colhead{(ks)}
		& \colhead{(UT)}
		& \colhead{(d)}
	}
	\startdata
	XMM\,E1 & 0882591901\tablenotemark{\scriptsize a} & 23.0 &  2022 December 26 & 249.2\\
	XMM\,E2 & 0915390201\tablenotemark{\scriptsize b} & 75.5 &  2023 January 05 & 259.7\\
	XMM\,E3 & 0932390701\tablenotemark{\scriptsize c} & 43.0 &  2023 December 24  & 601.0\\
	\enddata
	\tablenotetext{\scriptsize a}{GO program PI: S. Gezari.}
	\tablenotetext{\scriptsize b}{DDT request submitted by M. Guolo.}
	\tablenotetext{\scriptsize c}{DDT request submitted by Y. Yao.}
\end{deluxetable}

\section{Details of X-ray Analysis}

\subsection{X-ray Environment} \label{subsec:Xray_image}

\begin{figure}[htbp!]
    \centering 
    \includegraphics[width=0.45\columnwidth]{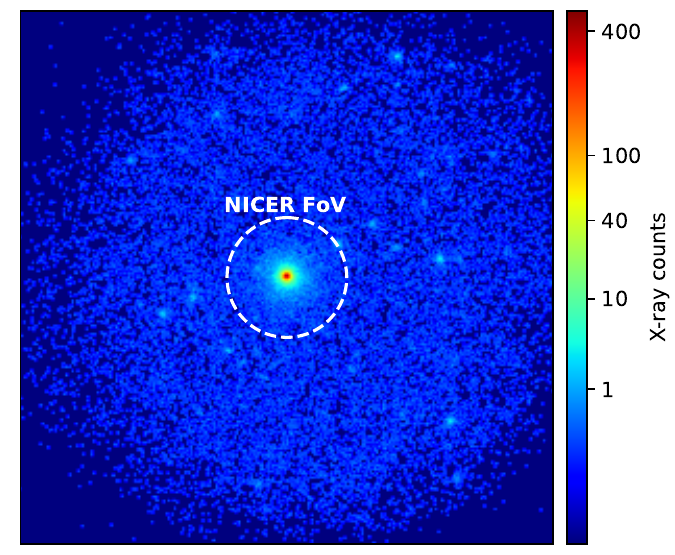}
    \caption{Stacked XRT (0.3--10\,keV) image centered around AT2022lri. The \nicer FoV is shown by the dashed circle with a radius of 3.1\,arcmin.
    \label{fig:XRT_stacked_image}}
\end{figure}

We created a stacked XRT image of AT2022lri using all available XRT observations. Figure~\ref{fig:XRT_stacked_image} shows that AT2022lri is the only bright X-ray object within the \nicer FoV.

\subsection{Sanity Checks of Physically Motivated Models} \label{subsec:sanity_check}

We performed a few sanity checks to validate the physically motivated models adopted in \S\ref{subsubsec:dip}.

First, in model (C), to explore the possibility of a lower turbulent velocity, we created model grids (\texttt{xstar} and \texttt{emi}) with $v_{\rm turb}=10^3\,{\rm km\,s^{-1}}$ and conducted the same modeling processes. We found that the original model grids with $v_{\rm turb}=10^4\,{\rm km\,s^{-1}}$ provide statistically better fits.

Next, the ionization parameter $\xi \equiv L / (nR^2)$, where $L$ is the 
luminosity of the incident radiation, $n$ is the gas density, and $R$ is the distance from the radiation source \citep{Kallman2001}. 
Therefore, under the condition where $n$ and $R$ do not change during the rapid X-ray variability, one might expect the ionization parameter to be correlated with the incident luminosity. 
To investigate this scenario in the absorption/emission modeling, we compute the 1--1000\,Ryd \texttt{diskbb} X-ray luminosity in the best-fit model (C), corrected for both Galactic absorption and absorption from the ionized absorber.
We refitted the model to the seven observations by fixing the log$\xi$ parameter based on log$L$, with the log$\xi$ value in sp6 set at 3.00. 
The new best-fit model yields $\chi^2/dof = 109.63/98$, with best-fit spectral parameters closely matching those in Table~\ref{tab:nicer_drop1_modC1}. Specifically, the outflow exhibits $v_{\rm out}/c=-0.124$ (in \texttt{xstar}), $v_{\rm out}/c=-0.155$ (in \texttt{emi}), and $N_{\rm H}\in [0.59, 9.88]\times 10^{21}\,{\rm cm^{-2}}$.
Similarly, for the reflection modeling, we compute the 1--1000\,Ryd \texttt{bbody} X-ray luminosity in the best-fit model (D), corrected for Galactic absorption. We refitted the model to the seven observations, and fixed the log$\xi$ parameter based on log$L$, with the log$\xi$ value in sp2 fixed at 2.14. 
The new best-fit model gives $\chi^2/dof = 112.80/105$, with best-fit spectral parameters very similar to what is shown in Table~\ref{tab:nicer_drop1_modD}.
Specifically, the outflow exhibits ${\rm log}(n/{\rm cm^{-3}})=18.19$ and $v_{\rm out}/c=-0.342$. 
We conclude that our spectral modeling results do not sensitively depend on small variations in the log$\xi$ parameter.

Finally, since the system is inferred to be close to face-on (\S\ref{subsubsec:basic_consideration}), we refitted the seven observations by imposing $i=18.20^{\circ}$ (instead of $i=45^{\circ}$) in model (D), where $18.20^{\circ}$ is the smallest $i$ value in the model grid of \texttt{xillverTDE}. The best-fit model has $\chi^2/dof = 111.61/104$, with best-fit parameters closely matching those shown in Table~\ref{tab:nicer_drop1_modD}. 
This verifies that the reflection modeling result is not sensitive to the assumed inclination parameter.

\bibliography{main}{}
\bibliographystyle{aasjournal}

\end{document}